\documentclass[aps,prl,twocolumn,amsmath,amssymb]{revtex4}
\usepackage{stmaryrd}
\usepackage[utf8]{inputenc}
\usepackage{times}
\usepackage{graphicx}
\usepackage{color}
\usepackage{bm,bbold}
\usepackage{bbm}
\usepackage{enumerate}
\usepackage{color}
\graphicspath{figures} 
\usepackage{amsfonts, amsmath, amsthm, amssymb} 
\usepackage{mathtools}
\usepackage{dsfont}
\usepackage{array}
\usepackage[colorlinks,bookmarks=false,citecolor=blue,linkcolor=red,urlcolor=blue]{hyperref}

\begin{document}

\title{Detecting many-body Bell non-locality by solving Ising models}

\author{Ir\'en\'ee Fr\'erot}
\email{irenee.frerot@icfo.eu}
\affiliation{ICFO-Institut de Ciencies Fotoniques, The Barcelona Institute of Science and Technology, Av. Carl Friedrich Gauss 3, 08860 Barcelona, Spain
}
\affiliation{Max-Planck-Institut f{\"u}r Quantenoptik, D-85748 Garching, Germany}
\author{Tommaso Roscilde}
\email{tommaso.roscilde@ens-lyon.fr}
\affiliation{Laboratoire de Physique, CNRS UMR 5672, Ecole Normale Sup\'erieure de Lyon, Universit\'e de Lyon, 46 All\'ee d'Italie, Lyon, F-69364, France
}

\begin{abstract}
Bell non-locality represents the ultimate consequence of quantum entanglement, fundamentally undermining the classical tenet that spatially-separated degrees of freedom possess objective attributes independently of the act of their measurement. 
Despite its importance, probing Bell non-locality in many-body systems is considered to be a formidable challenge, with a computational cost scaling exponentially with system size. 
Here we propose and validate an efficient variational scheme, based on the solution of inverse Ising classical problems, which in polynomial time can probe whether an arbitrary set of quantum data is compatible with a local theory; and, if not, it delivers a many-body Bell inequality violated by the quantum data. We use our approach to unveil new many-body Bell inequalities, violated by suitable measurements on paradigmatic quantum states (the low-energy states of Heisenberg antiferromagnets), paving the way to systematic Bell tests in the many-body realm.
\end{abstract}

\maketitle


\emph{Introduction: Bell tests and quantum certification.}  Quantum correlations, such as entanglement \cite{Horodecki2009}, Einstein-Podolsky-Rosen correlations \cite{Uolaetal2019,Reidetal2009} and Bell correlations \cite{Brunneretal2014,Scaranibook}, are common features of microscopic ensembles of quantum degrees of freedom (d.o.f), such as the electronic and nuclear spins in atoms and molecules \cite{Dehollainetal2016}, or pairs of photons produced by parametric down-conversion \cite{Aspectetal1982}. Their persistence in many-body systems is a central issue: an obstruction to the scalability of quantum correlations would be the core feature of a putative quantum-to-classical transition \cite{Zurek1991,Schlosshauer2019}; and, in parallel, they are the essential resource for most quantum technologies of second generation \cite{Acinetal2018}.  
   In view of all this, the robust certification of quantum correlations in many-body systems stands as a central problem for theoretical as well as experimental quantum physics. 

  \begin{figure*}
\includegraphics[width=\linewidth]{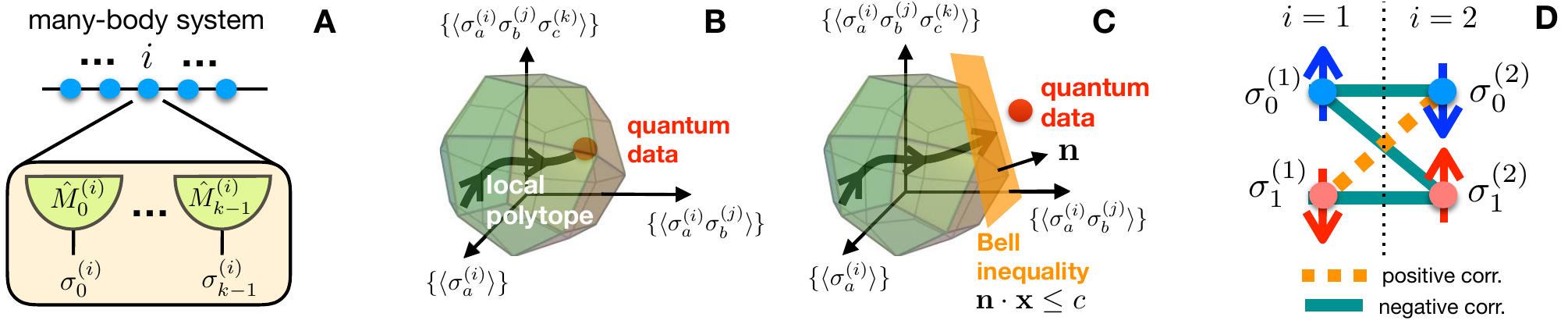}
\caption{{\bf Variational search of local-variable models.} (A) Sketch of the generic $(N,k,p)$ setting for Bell tests: each d.o.f of a quantum many-body system is subject to the measurement of $k$ different operators $\hat{M}_a$, with $p$ different outcomes $\sigma_a$ for each measurements; (B) Our Bell test of a set of quantum data -- comprising arbitrary moments ($\langle \sigma_a^{(i)}\rangle$, $\langle \sigma_a^{(i)} \sigma_b^{(j)} \rangle$, etc.) of the statistics of the measurement outcomes -- consists of generating a family of local-variable (LV) models which approximate the quantum data at best, describing a trajectory (black line) within the local polytope bounding all predictions of LV models; (C) if the LV predictions maintain a finite distance from the quantum data, this reveals the existence of a Bell inequality (corresponding to the closest polytope facet) which the quantum data violate;  (D) Frustrated correlation pattern among local variables in a (2,2,2) setting, which an LV model should reproduce in order to realize the correlations of a Bell pair $\left ( |\uparrow_1 \downarrow_2 \rangle - |\downarrow_1 \uparrow_2 \rangle \right )/\sqrt{2}$.}
\label{Fig1}
\end{figure*} 
   
   The most robust certification scheme is undoubtedly offered by the \emph{device-independent} (DI) approach, relying on the violation of a Bell inequality -- a so-called Bell test -- which does not assume anything about the quantum system except what can be assessed experimentally. Specifically, we assume that a many-body system is composed of $N$ spatially-separated d.o.f -- that we imagine as arranged over a lattice -- on which $k$ different observables (\emph{inputs}) can be experimentally measured; and each of them can deliver $p$ results (\emph{outputs}). We indicate with $\sigma_a^{(i)}$ the $p$ possible values of the $a$-th observable ($a=0,...,k-1$) on the $i$-th d.o.f ($i=1,...,N$) 
   -- these choices define a $(N,k,p)$ scenario for a Bell test (Fig.~\ref{Fig1}A). 
   Moreover, we indicate with $\langle f({\bm \sigma}) \rangle_{\rm Q}$ (where $\bm \sigma = \{\sigma_a^{(i)}\}$) the average value of any function $f$ of the measurement outputs -- hereafter denoted as quantum data. The DI approach certifies the strongest form of quantum correlations -- Bell non-locality -- when the quantum data violate a Bell inequality \cite{Brunneretal2014,Scaranibook}, a constrain for all local-variable (LV) models, designed to capture the most general form of classical correlations. First envisioned by Bell \cite{Bell1964}, such models are defined by a joint probability distribution $P_{\rm LV}(\bm \sigma)$ for all measurement outcomes, treated as classical variables \cite{Fine1982}. 
 If the dataset involves the outcomes of incompatible measurements, such a joint probability distribution is not admitted by quantum mechanics, creating a fundamental tension with LV models. In the following we indicate with $\langle ... \rangle_{\rm LV}$ an average over the $P_{\rm LV}$ distribution. The simplest Bell inequalities are linear combinations of few-body expectation values:
\begin{equation}
 \sum_{i=1}^N \sum_{a=1}^k  \alpha_a^{(i)} \langle \sigma_a^{(i)} \rangle_{\rm LV} + \sum_{i <j} \sum_{a,b=1}^{k} \beta_{a,b}^{(i,j)} \langle \sigma_a^{(i)} \sigma_b^{(j)} \rangle_{\rm LV} + ... \geq -B_c
 \label{e.Bell}
 \end{equation}  
 where $-B_c$ is the so-called classical bound. Geometrically, every such inequality defines a hyperplane in the space of correlations, separating two half-spaces, one of which contains all datasets compatible with LV models. The intersection of these half-spaces defines the so-called local polytope (Fig.~\ref{Fig1}B). Certifying Bell non-locality corresponds then to assessing that the quantum data of interest lie outside the local polytope (Fig.~\ref{Fig1}C).

    \emph{The quantum membership problem.} 
  The search for Bell inequalities violated by quantum many-body data for systems with $N \gg 1$ represents a formidable task. Indeed, given a quantum dataset $\{ \langle f_r(\bm \sigma) \rangle_{\rm Q}; r = 1,..,R \}$  -- where the $f_r({\bm \sigma})$'s are terms such as $\sigma_a^{(i)}$ or
$\sigma_a^{(i)} \sigma_b^{(j)}$ in Eq.~\eqref{e.Bell} --
  the local polytope has $p^{kN}$ vertices, and its full reconstruction has a prohibitive (exponential) cost \cite{Brunneretal2014}. Many-body Bell inequalities have been successfully identified in the past \cite{Mermin1990,Guehneetal2005,tura_energy_2017}, but they are violated only by selected quantum states \cite{Lanyonetal2014}. More systematic strategies have been devised recently that either restrict the search to Bell inequalities which are fully symmetric under exchange of lattice-site indices (namely with $\alpha_a^{(i)} = \alpha_a$, $\beta_{a,b}^{(i,j)} = \beta_{a,b}$, etc. in Eq.~\eqref{e.Bell}) \cite{Turaetal2014,fadelT2017}, or to inequalities which only involve a restricted range of correlations under translational invariance \cite{wangN2017}, circumventing the exponential cost but losing in generality; an alternative strategy is that of approximating the local polytope from the outside \cite{Baccarietal2017}, with an exponential cost for the approximation to converge to the exact polytope. Hence the \emph{quantum membership problem} (``Does a set of quantum data belong to the local polytope?") is considered to be an exponentially hard one. Our main result is to exhibit an algorithm solving this problem in polynomial time under very general assumptions; and to validate such an approach by discovering new Bell inequalities violated by relevant quantum many-body states in the thermodynamic limit. 
  
    \emph{Solving the membership problem by inverse statistical methods.} Our approach to the above problem consists of trying to explicitly build an LV model $P_{\rm LV}$ which  reproduces the quantum data, namely such that $\langle f_r(\bm \sigma) \rangle_{\rm LV} = \langle f_r(\bm \sigma) \rangle_{\rm Q}$ for all $r=1,...,R$ (within the error bar of the quantum data). In a realistic scenario, $R$ scales polynomially with $N$; therefore, if such a distribution exists, it is certainly not unique, because it can be parametrized by many more parameters ($p^{kN}-1$ independent values) than the number $R$ of constraints. Yet, if multiple distributions exist, there is one of them which is least biased, parametrized by the \emph{minimal} number of parameters. This distribution maximizes Shannon entropy under the constraints \cite{Jaynes1957, Tikochinskyetal1984}, or equivalently minimizes the ``free-energy" functional 
   \begin{equation}
   F[P_{\rm LV}] = \sum_{\bm \sigma} P_{\rm LV} \log P_{\rm LV} - \sum_r K_r \left ( \langle f_r \rangle_{\rm LV} - \langle f_r \rangle_{\rm Q} \right )~.
   \end{equation} 
   The solution takes the form of a Boltzmann distribution \cite{Jaynes1957}
  \begin{equation}
  P_{\rm LV}(\bm \sigma) = \exp[\sum_r K_r  f_r(\bm \sigma)]/{\cal Z}
  \label{e.Boltzmann}
  \end{equation} 
  in which the Lagrange multipliers $K_r$ (forming the vector ${\bm K}=\{ K_r \}$) play the role of coupling constants defining an effective Hamiltonian ${\cal H}(\bm \sigma;\bm K)= -\sum_r K_r  f_r(\bm \sigma)$, and ${\cal Z}$ is the corresponding partition function. Therefore, our central observation is the following: if a LV model reproducing the quantum data exists, it can be found in the form of Eq.~\eqref{e.Boltzmann} upon adjusting the coupling constants.  In the case of binary outcomes ($p=2$), to which we hereafter specialize, the $\sigma$'s are classical Ising variables ($\sigma=\pm 1$), and therefore the LV model represents the equilibrium Boltzmann distribution of a generalized classical Ising model with Hamiltonian ${\cal H}$. 

   In summary, without loss of generality, the problem is reduced to adjusting the coupling constants of a classical Ising model so as to fit the quantum data. This, however, is a well-known problem in statistical inference, namely an \emph{inverse Ising problem} \cite{Nguyenetal2017}, which has the remarkable feature of being a convex optimization problem upon introducing the following cost function
   \begin{equation} 
   {\cal L}(\bm K) =  \log {\cal Z}({\bm K})  - \sum_r K_r \langle f_r\rangle_{\rm Q}
   \end{equation} 
   where ${\cal L}$ is related to (minus) the log-likelihood. Indeed, the Hessian of the cost function 
   \begin{equation} 
   H_{rs}  =\frac{ \partial^2 {\cal L}}{\partial K_r \partial K_s} = \langle f_r f_s \rangle_{\rm LV} -  \langle f_r \rangle_{\rm LV} \langle f_s \rangle_{\rm LV}
   \end{equation}
    is the covariance matrix of the $f_r$ functions, and is therefore semi-definite positive. The convexity of the cost function implies that a simple gradient-descent algorithm, following the gradient ${\bm G} = \{ G_r \}$ of the cost function:
  \begin{equation}
   G_r = \frac{\partial{\cal L}}{\partial K_r} = \langle f_r \rangle_{\rm LV} - \langle f_r \rangle_{\rm Q} ~,
   \label{e.gradient}
\end{equation}
is guaranteed to converge to the global minimum \cite{Boyd2004convex}.

\emph{Building a data-tailored Bell inequality.}
 Our algorithm presents then two possible behaviors: 1) if the quantum data are reproducible by an LV model, it converges to well-defined couplings ${\bm K}$ which lead to the vanishing of the gradient ${\bm G}$ [Eq.~\eqref{e.gradient}], namely of the distance vector between the quantum data and the LV predictions (Fig.~\ref{Fig1}B) ; 2) otherwise, the quantum data lie outside of the local polytope, so that ${\bm G}$ remains necessarily finite, leading to a runaway to infinity of the coupling constants as updated by the gradient-descent algorithm: $K_r' = K_r - \epsilon G_r$ (with $\epsilon\ll 1$ the step variable in the numerical implementation of the gradient descent). In this case, the algorithm converges in practice when the minimal distance $|{\bm G}_{\infty}|^2 = \min_{\rm LV} \sum_r \langle (f_r \rangle_{\rm LV} - \langle f_r \rangle_{\rm Q} )^2$ between the LV predictions and the classical data is attained numerically. This convergence criterion marks the fact that the variational search of the LV model has hit from the inside a facet of the polytope (Fig.~\ref{Fig1}C), defining a Bell inequality violated by the quantum data. The latter inequality stems from a simple rewriting of the condition $|{\bm G}_{\infty}|^2 >0$, namely
 %
 \begin{equation}
  \sum_r  G_{r,\infty} \langle f_r \rangle_{\rm Q} < \min_{\rm LV} \sum_r G_{r,\infty} \langle f_r \rangle_{\rm LV} = -B_c ~.
 \label{e.emergentBell}
 \end{equation} 
 
The minimization of the right-hand side of Eq.~\eqref{e.emergentBell} -- defining the classical bound $B_c$ -- is attained as the ground-state energy of the classical Hamiltonian ${\cal K}$ (not to be confused with $\cal H$): ${\cal K}(\bm \sigma) = \sum_r G_{r, \infty} f_r({\bm \sigma})$.
Interestingly, we observe that ${\cal K}(\bm \sigma)$ is necessarily a frustrated Hamiltonian, namely a function whose minimum is not obtained by minimizing each term $G_{r,\infty} f_r({\bm \sigma})$ individually. Indeed, in the absence of frustration, the quantum data has no chance of being strictly lower than the classical bound defined in Eq.~\eqref{e.emergentBell}.

   \begin{figure}
\includegraphics[width=\linewidth]{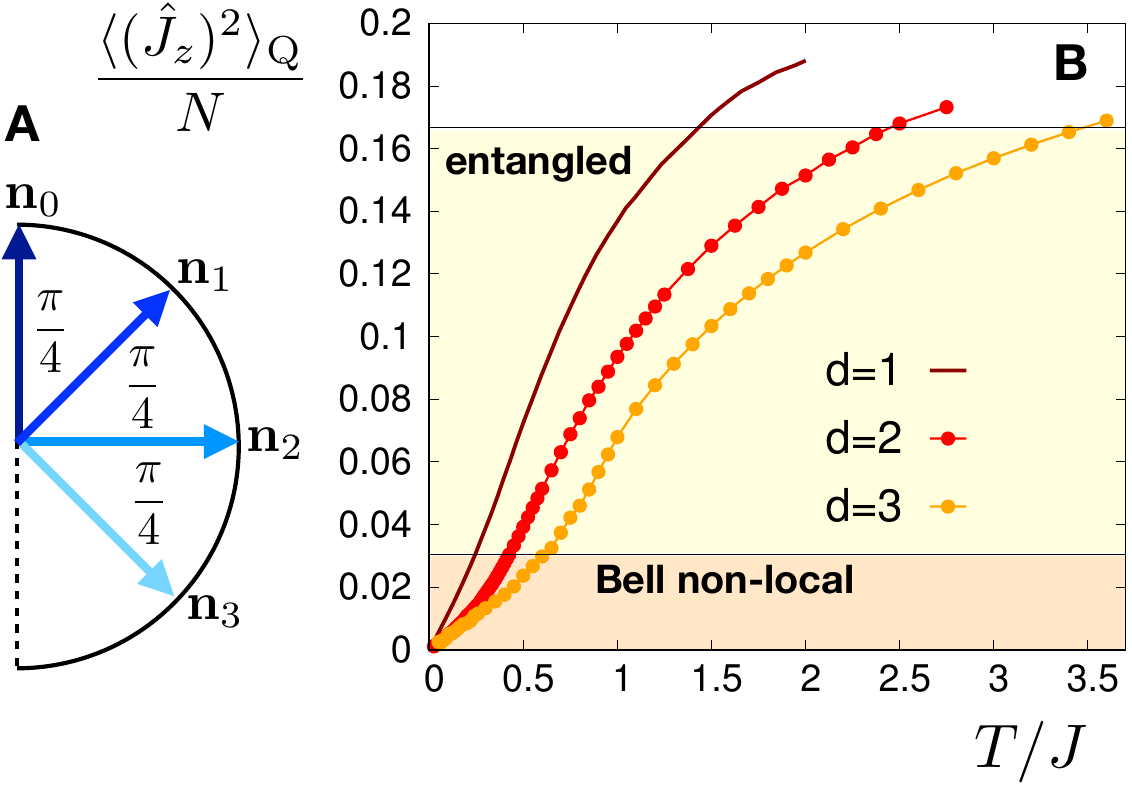}
\caption{{\bf Many-body Bell non-locality of quantum Heisenberg antiferromagnets.}  (A) Measurement basis ($k=4$) providing the strongest violation of the Bell inequality in Eq.~\eqref{e.mbPBC} by the low-temperature data of quantum Heisenberg antiferromagnets; (B) Normalized fluctuations of a collective spin component, $\langle (\hat{J}^z)^2 \rangle/N$, in Heisenberg antiferromagnets on the linear chain ($d=1$), the square lattice ($d=2$) and the cubic lattice ($d=3$). The data shown are obtained via the Bethe-Ansatz prediction \cite{Eggertetal1994} for $d=1$, and by quantum Monte Carlo ($d=2,3$) on lattices of size $30^2$ and $12^3$ respectively -- the thermodynamic limit values are essentially reached for these sizes. When the fluctuations become smaller than the classical bound $\beta_4$ (see text) they witness the appearance of Bell non-locality; in the figure we report as well the known bound for witnessing entanglement \cite{Toth2004,Wiesniaketal2005}.}     
\label{Fig2}
\end{figure} 

 Before demonstrating the practical use of our approach, we would like to point out its computational efficiency. Its strength relies fundamentally upon its data-driven nature: instead of trying to reconstruct the whole local polytope (potentially producing a large number of unviolated Bell inequalities), it directly tests for the non-locality of a particular dataset; and it delivers a Bell inequality violated by the available quantum data. Its main computational cost is imposed by the calculation of the statistical averages $\langle f_r({\bm \sigma})\rangle_{\rm LV}$: such a calculation is generically efficient and scalable to arbitrary $N$ by using classical Monte Carlo, unless the classical Ising Hamiltonian ${\cal H}(\bm \sigma;\bm K)$ happens to be a spin-glass model -- something which is categorically avoided if the quantum data have elementary spatial symmetries, and if the local observables are not chosen randomly. Otherwise, the computational cost to reach a relative precision of $\varepsilon$ scales at worst as ${\cal O}(N^{n+z/d} \times \varepsilon^{-2})$ if the $f_r$'s are correlation functions involving up to $n$ points -- here $z$ is the dynamical critical exponent, which is non-zero (and ${\cal O}(1)$) only if the classical Ising model sits exactly at a critical point (see Supplemental Material - SM - for further discussion \cite{SM}).
 
As explained above, our approach starts from a thoughtfully chosen set of quantum data: in the following we illustrate it in three paradigmatic cases, in which the input quantum data are offered by the spin expectation values and 2-point correlation functions of 1) a Bell pair; 2) the quantum critical point of the $d=2$ transverse-field Ising model; and 3) the low-energy states of the Heisenberg antiferromagnet on hyper-cubic lattices.

   \emph{Bell pair: failure of LV theories from frustration.} We first explain the conceptual significance of our approach in the paradigmatic case of a Bell pair $(|\uparrow_1 \downarrow_2\rangle - |\downarrow_1 \uparrow_2\rangle)/\sqrt{2}$  of two $S=1/2$ spins. In the case of a (2,2,2) scenario, choosing the measurements $\hat \sigma_0^{(1)} = \hat \sigma_x$,
$\hat \sigma_1^{(1)} = \hat \sigma_y$;  $\hat \sigma_0^{(2)} = \cos\theta ~\hat \sigma_x - \sin \theta ~\hat \sigma_y$, $\hat \sigma_1^{(2)} = \cos\theta ~\hat\sigma_x + \sin \theta~ \hat\sigma_y$, one obtains the following quantum data for the correlation functions, $\langle \sigma_0^{(1)} \sigma_0^{(2)} \rangle_{\rm Q} = \langle \sigma_1^{(1)} \sigma_1^{(2)} \rangle_{\rm Q} = -\cos\theta$, and $\langle \sigma_0^{(1)} \sigma_1^{(2)} \rangle_{\rm Q} = -\langle \sigma_1^{(1)} \sigma_0^{(2)} \rangle_{\rm Q} = -\sin\theta$. Notice that $\hat \sigma^{(i)}_a$'s are quantum operators, while $\sigma^{(i)}_a$'s  are classical Ising variables representing the binary outcomes of their measurement. Choosing the optimal angle $\theta = \pi/4$ leads to correlation functions which take the common absolute value $1/\sqrt{2}$, but realize a fully frustrated correlation loop (three negative correlations and a positive one -- see Fig.~\ref{Fig1}D). When trying to reproduce this correlation pattern with the equilibrium state of a classical Ising model ${\cal H} = -\sum_{a,b\in\{0,1\}} K_{ab} ~\sigma_a^{(1)} \sigma_b^{(2)}$, one can easily realize that the optimal choice is to take $K_{ab} = \beta J_{ab}$ with $\beta \to \infty$ (restricting the phase space to the ground state manifold of the Hamiltonian) and $J_{00} = J_{11} = J_{01} = - J_{10}$, defining a fully frustrated square (3 antiferromagnetic couplings and a ferromagnetic one), such that $\langle \sigma_0^{(1)} \sigma_0^{(2)} \rangle_{\rm LV} =  \langle \sigma_1^{(1)} \sigma_1^{(2)} \rangle_{\rm LV} = \langle \sigma_0^{(1)} \sigma_1^{(2)} \rangle_{\rm LV} = - \langle \sigma_1^{(1)} \sigma_0^{(2)} \rangle_{\rm LV} = -1/4$ (since $\cal H$ has 8 degenerate ground states, in which there is always one correlation function out of 4 with the wrong sign).  As a consequence, one obtains 
for the gradient vector the components $G_{00} = G_{11} = G_{01} = -G_{10} = (2\sqrt{2}-1)/4$ defining an effective Hamiltonian ${\cal K}$ which has the same form as $\cal H$, and which reconstructs the celebrated 
Clauser-Horne-Shimony-Holt (CHSH) inequality \cite{Clauseretal1969}   $\langle \sigma^{(1)}_0 \sigma^{(2)}_0 + \sigma^{(1)}_1 \sigma^{(2)}_1 + \sigma^{(1)}_0 \sigma^{(2)}_1 - \sigma^{(1)}_0 \sigma^{(2)}_1 \rangle_{\rm LV} \geq -B_c = -2$ (while the quantum data achieve the value $-2\sqrt{2}$). 

  \emph{Bell inequality for the quantum Ising model at its quantum-critical point.} Moving on to many-body systems, we consider the 2$d$ transverse-field Ising model at its quantum-critical (QC) point. Here, the quantum data consist of the net magnetization and pair correlation functions, and our approach reconstructs a permutationally-invariant Bell inequality violated by the quantum data \cite{SM}. The relevant inequality, first identified in Ref.~\cite{Turaetal2014}, is violated by strongly squeezed states \cite{Schmiedetal2016,Engelsenetal2017,Pigaetal2019}, and squeezing is also a property of the QC point in question \cite{FrerotR2018}. Yet our approach allows us to make a stronger statement, namely that -- given the $(N,2,2)$ scheme with measurement bases suggested by Ref.~\cite{Schmiedetal2016} --  a symmetric Bell inequality is the optimal one, namely the one which is most strongly violated, for quantum data that are not at all symmetric (unlike those produced in the experiments of Refs.~\cite{Schmiedetal2016,Engelsenetal2017}), because of the spatial decay of correlations functions at criticality. 
 
  \emph{Bell inequality for Heisenberg antiferromagnets.}  We conclude our article by focusing on the equilibrium states of a paradigmatic quantum spin-lattice model, namely the quantum Heisenberg antiferromagnet (QHAF) with Hamiltonian $\hat{\cal H} = J \sum_{\langle ij \rangle} \hat{\bm S}^{(i)} \cdot \hat{\bm S}^{(j)}$, where $\hat {\bm S}^{(i)}$ are quantum $S=1/2$ operators, and the sum runs over pairs of nearest neighbors on a hypercubic lattice with an even number of sites. The ground state of this model realizes a global singlet, namely a many-body generalization of the Bell pair considered above. 
 We focus on a $(N,k,2)$ scenario, with $k\ge 3$ measurements per spin along axes ${\bm n}_a$ $(a=0,1,...,k-1)$, and we consider a uniform measurement strategy in which the axes are coplanar and form an angle $a\pi / k$ with a given reference axis (this turns out to be optimal  \cite{SM}) -- see Fig.~\ref{Fig2}A. Feeding our algorithm with the two-point correlation function of the 2$d$ QHAF with $k=3$ measurements as quantum data, we discover that the latter violate the following symmetric Bell inequality 
 \begin{equation}
\langle {\cal B} \rangle_{\rm LV} =  \sum_{a=0}^{k-1} {\cal S}_{aa} + \sum_{a=0}^{k-2} {\cal S}_{a,a+1} - {\cal S}_{k-1,0} \geq -B_c = -2N(k-1) ~,
\label{e.mbPBC}
 \end{equation}
where ${\cal S}_{ab} = \sum_{i\neq j} \langle \sigma_a^{(i)} \sigma_b^{(j)} \rangle_{\rm LV}$. This inequality (proven in the SM \cite{SM}) turns out to be a many-body extension of the Pearle-Braunstein-Caves inequality \cite{Pearle1970, BraunsteinC1990} proposed for non-locality detection in 2-spin states. 
Similarly to the above-cited example of the QC point of the 2$d$ quantum Ising model, it is remarkable to notice that quantum data with spatial structure -- such as the correlation function of the 2$d$QHAF -- are found to most strongly violate a Bell inequality in which the spatial structure is washed out by the symmetrization procedure.

To see explicitly that the ground state of the QHAF violates the inequality of Eq.~\eqref{e.mbPBC}, we make use of the SU(2) invariance to rewrite the Bell operator $\hat{\cal B}$  in the form \cite{SM}: $\hat{\cal B} = 4k[1+\cos(\pi/k)] \hat{J}_z^2 - Nk\cos(\pi/k) - Nk$, where $\hat J_z = \sum_i \hat{S}_z^{(i)}$ is the collective spin along $z$. Therefore, the classical bound $-B_c$ is violated by the quantum data whenever
\begin{equation}
\frac{\langle \hat{J}_z^2 \rangle_{\rm Q}}{N} < \beta_k =   \frac{1}{4} - \frac{k-1}{2k \left ( 1+ \cos \frac{\pi}{k} \right)}~.
\label{e.witness}
\end{equation}
where the largest value of the right-hand side is found for $k=4$, and it reads $\beta_4 = 1/(16+12\sqrt{2}) = 0.030330...$ ~.  Eq.~\eqref{e.witness} states that a sufficiently low value of the variance of one collective spin component (below the $\beta_4$ bound) is a witness \cite{Schmiedetal2016} of Bell non-locality (``witness" because, in order to derive it, we explicitly used the spin algebra as well as the hypothesis of SU(2) invariance of the state). The ground state of all Heisenberg antiferromagnets with even $N$ (regardless of the geometry of the underlying lattice) are total spin singlets (such that $\langle \hat{J}_z^2 \rangle_{\rm Q}=0$), and hence they satisfy the above criterion and violate the Bell inequality of Eq.~\eqref{e.mbPBC}. Moreover, in the SM \cite{SM} we show that the quantum violation of the inequality Eq.~\eqref{e.mbPBC} offered by total spin singlets, $\langle \hat{\cal B} \rangle_{\rm Q} = -Nk[1 + \cos(\pi /k)]$, is the maximal violation authorized by quantum mechanics (namely, regardless of the dimension of the Hilbert space of the system, of its quantum state, and of the chosen measurements).  
Fig.~\ref{Fig2}B shows that the condition Eq.~\eqref{e.witness} is also met by thermal equilibrium states of the QHAF in $d=$1, 2 and 3 up to very sizable temperatures (the higher the larger $d$ is, as non-locality is clearly protected by the strength of antiferromagnetic correlations). The condition of Eq.~\eqref{e.witness} is to be contrasted with the much looser one, $\langle \hat{J}_z^2 \rangle_{\rm Q}/N < 1/6$ required to witness entanglement between the individual spins \cite{Toth2004,Wiesniaketal2005} -- namely to exclude the possibility of writing the state of the system as $\hat\rho = \sum_s p_s \otimes_i \hat\rho_s^{(i)}$, where $\hat\rho_s^{(i)}$ are arbitrary (pure or mixed) states of individual spins. This 
reflects the fact that, for mixed states, Bell non-locality is a much stronger form of quantum correlations than entanglement. Moreover the fundamental connection between the collective spin variance and the spin susceptibility at thermal equilibrium 
$\chi_z = \langle \hat{J}_z^2 \rangle_{\rm Q}/ (k_B T N)$ (where $T$ is the temperature) makes the above witness of non-locality experimentally accessible to magnetometry experiments on quantum magnets at realistic temperatures. 

\emph{Conclusions.} We have demonstrated a variational approach which can assess whether an arbitrary set of quantum data, coming from scalable many-body systems, exhibits quantum non-locality; and which reconstructs the Bell inequality most strongly violated by the data at hand. The computational cost of the algorithm is polynomial in system size whenever the quantum data are not obtained from systems governed by random Hamiltonians, and are not obtained by using a random measurement basis for each d.o.f -- and it may still remain polynomial even if the above conditions are not met. 
 Therefore, our approach opens the door to scalable and systematic certification of entanglement in synthetic quantum matter (quantum simulators \cite{Georgescuetal2013,Altmanetal2019}, quantum processors \cite{Wendin2017,Bruzewiczetal2019}). When the violated Bell inequalities have a symmetric structure under the exchange of d.o.f (as in the case of the Heisenberg antiferromagnets reported in this work), a witness of Bell non-locality can be formulated by using collective observables only \cite{Schmiedetal2016}, and the latter is therefore accessible in the broader context of quantum materials in condensed matter physics. 

{\emph{Acknowledgments} We warmly thank A. Ac{\'i}n, A. Aloy, A. Aspect, F. Baccari, C. Branciard, M. Lewenstein, P. Ronceray, L. Tagliacozzo, and J. Tura for insightful discussions and encouragement. This work is supported by ANR (``EELS'' project), QuantERA (``MAQS'' project), the Spanish MINECO (Severo Ochoa SEV-2015-0522), the fundacio Mir-Puig and Cellex through an ICFO-MPQ postdoctoral fellowship, and the Generalitat de Catalunya (SGR 1381, QuantumCAT and CERCA Programme).


\bibliography{biblio_variational_nonlocality}

\begin{thebibliography}{50}
\expandafter\ifx\csname natexlab\endcsname\relax\def\natexlab#1{#1}\fi
\expandafter\ifx\csname bibnamefont\endcsname\relax
  \def\bibnamefont#1{#1}\fi
\expandafter\ifx\csname bibfnamefont\endcsname\relax
  \def\bibfnamefont#1{#1}\fi
\expandafter\ifx\csname citenamefont\endcsname\relax
  \def\citenamefont#1{#1}\fi
\expandafter\ifx\csname url\endcsname\relax
  \def\url#1{\texttt{#1}}\fi
\expandafter\ifx\csname urlprefix\endcsname\relax\def\urlprefix{URL }\fi
\providecommand{\bibinfo}[2]{#2}
\providecommand{\eprint}[2][]{\url{#2}}

\bibitem[{\citenamefont{Horodecki et~al.}(2009)\citenamefont{Horodecki,
  Horodecki, Horodecki, and Horodecki}}]{Horodecki2009}
\bibinfo{author}{\bibfnamefont{R.}~\bibnamefont{Horodecki}},
  \bibinfo{author}{\bibfnamefont{P.}~\bibnamefont{Horodecki}},
  \bibinfo{author}{\bibfnamefont{M.}~\bibnamefont{Horodecki}},
  \bibnamefont{and}
  \bibinfo{author}{\bibfnamefont{K.}~\bibnamefont{Horodecki}},
  \bibinfo{journal}{Rev. Mod. Phys.} \textbf{\bibinfo{volume}{81}},
  \bibinfo{pages}{865} (\bibinfo{year}{2009}),
  \urlprefix\url{https://link.aps.org/doi/10.1103/RevModPhys.81.865}.

\bibitem[{\citenamefont{{Uola} et~al.}(2019)\citenamefont{{Uola}, {Costa},
  {Chau Nguyen}, and {G{\"u}hne}}}]{Uolaetal2019}
\bibinfo{author}{\bibfnamefont{R.}~\bibnamefont{{Uola}}},
  \bibinfo{author}{\bibfnamefont{A.~C.~S.} \bibnamefont{{Costa}}},
  \bibinfo{author}{\bibfnamefont{H.}~\bibnamefont{{Chau Nguyen}}},
  \bibnamefont{and}
  \bibinfo{author}{\bibfnamefont{O.}~\bibnamefont{{G{\"u}hne}}},
  \bibinfo{journal}{arXiv e-prints} \bibinfo{eid}{arXiv:1903.06663}
  (\bibinfo{year}{2019}), \eprint{1903.06663}.

\bibitem[{\citenamefont{Reid et~al.}(2009)\citenamefont{Reid, Drummond, Bowen,
  Cavalcanti, Lam, Bachor, Andersen, and Leuchs}}]{Reidetal2009}
\bibinfo{author}{\bibfnamefont{M.~D.} \bibnamefont{Reid}},
  \bibinfo{author}{\bibfnamefont{P.~D.} \bibnamefont{Drummond}},
  \bibinfo{author}{\bibfnamefont{W.~P.} \bibnamefont{Bowen}},
  \bibinfo{author}{\bibfnamefont{E.~G.} \bibnamefont{Cavalcanti}},
  \bibinfo{author}{\bibfnamefont{P.~K.} \bibnamefont{Lam}},
  \bibinfo{author}{\bibfnamefont{H.~A.} \bibnamefont{Bachor}},
  \bibinfo{author}{\bibfnamefont{U.~L.} \bibnamefont{Andersen}},
  \bibnamefont{and} \bibinfo{author}{\bibfnamefont{G.}~\bibnamefont{Leuchs}},
  \bibinfo{journal}{Rev. Mod. Phys.} \textbf{\bibinfo{volume}{81}},
  \bibinfo{pages}{1727} (\bibinfo{year}{2009}),
  \urlprefix\url{https://link.aps.org/doi/10.1103/RevModPhys.81.1727}.

\bibitem[{\citenamefont{Brunner et~al.}(2014)\citenamefont{Brunner, Cavalcanti,
  Pironio, Scarani, and Wehner}}]{Brunneretal2014}
\bibinfo{author}{\bibfnamefont{N.}~\bibnamefont{Brunner}},
  \bibinfo{author}{\bibfnamefont{D.}~\bibnamefont{Cavalcanti}},
  \bibinfo{author}{\bibfnamefont{S.}~\bibnamefont{Pironio}},
  \bibinfo{author}{\bibfnamefont{V.}~\bibnamefont{Scarani}}, \bibnamefont{and}
  \bibinfo{author}{\bibfnamefont{S.}~\bibnamefont{Wehner}},
  \bibinfo{journal}{Rev. Mod. Phys.} \textbf{\bibinfo{volume}{86}},
  \bibinfo{pages}{419} (\bibinfo{year}{2014}),
  \urlprefix\url{https://link.aps.org/doi/10.1103/RevModPhys.86.419}.

\bibitem[{\citenamefont{Scarani}(2019)}]{Scaranibook}
\bibinfo{author}{\bibfnamefont{V.}~\bibnamefont{Scarani}},
  \emph{\bibinfo{title}{Bell nonlocality}} (\bibinfo{publisher}{Oxford
  University Press}, \bibinfo{year}{2019}).

\bibitem[{\citenamefont{Dehollain et~al.}(2016)\citenamefont{Dehollain,
  Simmons, Muhonen, Kalra, Laucht, Hudson, Itoh, Jamieson, McCallum, Dzurak
  et~al.}}]{Dehollainetal2016}
\bibinfo{author}{\bibfnamefont{J.~P.} \bibnamefont{Dehollain}},
  \bibinfo{author}{\bibfnamefont{S.}~\bibnamefont{Simmons}},
  \bibinfo{author}{\bibfnamefont{J.~T.} \bibnamefont{Muhonen}},
  \bibinfo{author}{\bibfnamefont{R.}~\bibnamefont{Kalra}},
  \bibinfo{author}{\bibfnamefont{A.}~\bibnamefont{Laucht}},
  \bibinfo{author}{\bibfnamefont{F.}~\bibnamefont{Hudson}},
  \bibinfo{author}{\bibfnamefont{K.~M.} \bibnamefont{Itoh}},
  \bibinfo{author}{\bibfnamefont{D.~N.} \bibnamefont{Jamieson}},
  \bibinfo{author}{\bibfnamefont{J.~C.} \bibnamefont{McCallum}},
  \bibinfo{author}{\bibfnamefont{A.~S.} \bibnamefont{Dzurak}},
  \bibnamefont{et~al.}, \bibinfo{journal}{Nature Nanotechnology}
  \textbf{\bibinfo{volume}{11}}, \bibinfo{pages}{242} (\bibinfo{year}{2016}),
  ISSN \bibinfo{issn}{1748-3395},
  \urlprefix\url{https://doi.org/10.1038/nnano.2015.262}.

\bibitem[{\citenamefont{Aspect et~al.}(1982)\citenamefont{Aspect, Grangier, and
  Roger}}]{Aspectetal1982}
\bibinfo{author}{\bibfnamefont{A.}~\bibnamefont{Aspect}},
  \bibinfo{author}{\bibfnamefont{P.}~\bibnamefont{Grangier}}, \bibnamefont{and}
  \bibinfo{author}{\bibfnamefont{G.}~\bibnamefont{Roger}},
  \bibinfo{journal}{Phys. Rev. Lett.} \textbf{\bibinfo{volume}{49}},
  \bibinfo{pages}{91} (\bibinfo{year}{1982}),
  \urlprefix\url{https://link.aps.org/doi/10.1103/PhysRevLett.49.91}.

\bibitem[{\citenamefont{Zurek}(1991)}]{Zurek1991}
\bibinfo{author}{\bibfnamefont{W.}~\bibnamefont{Zurek}},
  \bibinfo{journal}{Physics Today} \textbf{\bibinfo{volume}{44}},
  \bibinfo{pages}{36} (\bibinfo{year}{1991}).

\bibitem[{\citenamefont{Schlosshauer}(2019)}]{Schlosshauer2019}
\bibinfo{author}{\bibfnamefont{M.}~\bibnamefont{Schlosshauer}},
  \bibinfo{journal}{Physics Reports} \textbf{\bibinfo{volume}{831}},
  \bibinfo{pages}{1 } (\bibinfo{year}{2019}), ISSN \bibinfo{issn}{0370-1573},
  \bibinfo{note}{quantum decoherence},
  \urlprefix\url{http://www.sciencedirect.com/science/article/pii/S0370157319303084}.

\bibitem[{\citenamefont{Ac{\'i}n et~al.}(2018)\citenamefont{Ac{\'i}n, Bloch,
  Buhrman, Calarco, Eichler, Eisert, Esteve, Gisin, Glaser, Jelezko
  et~al.}}]{Acinetal2018}
\bibinfo{author}{\bibfnamefont{A.}~\bibnamefont{Ac{\'i}n}},
  \bibinfo{author}{\bibfnamefont{I.}~\bibnamefont{Bloch}},
  \bibinfo{author}{\bibfnamefont{H.}~\bibnamefont{Buhrman}},
  \bibinfo{author}{\bibfnamefont{T.}~\bibnamefont{Calarco}},
  \bibinfo{author}{\bibfnamefont{C.}~\bibnamefont{Eichler}},
  \bibinfo{author}{\bibfnamefont{J.}~\bibnamefont{Eisert}},
  \bibinfo{author}{\bibfnamefont{D.}~\bibnamefont{Esteve}},
  \bibinfo{author}{\bibfnamefont{N.}~\bibnamefont{Gisin}},
  \bibinfo{author}{\bibfnamefont{S.~J.} \bibnamefont{Glaser}},
  \bibinfo{author}{\bibfnamefont{F.}~\bibnamefont{Jelezko}},
  \bibnamefont{et~al.}, \bibinfo{journal}{New Journal of Physics}
  \textbf{\bibinfo{volume}{20}}, \bibinfo{pages}{080201}
  (\bibinfo{year}{2018}),
  \urlprefix\url{https://doi.org/10.1088%2F1367-2630%2Faad1ea}.

\bibitem[{\citenamefont{Bell}(1964)}]{Bell1964}
\bibinfo{author}{\bibfnamefont{J.~S.} \bibnamefont{Bell}},
  \bibinfo{journal}{Physics} \textbf{\bibinfo{volume}{1}}, \bibinfo{pages}{195}
  (\bibinfo{year}{1964}).

\bibitem[{\citenamefont{Fine}(1982)}]{Fine1982}
\bibinfo{author}{\bibfnamefont{A.}~\bibnamefont{Fine}}, \bibinfo{journal}{Phys.
  Rev. Lett.} \textbf{\bibinfo{volume}{48}}, \bibinfo{pages}{291}
  (\bibinfo{year}{1982}),
  \urlprefix\url{https://link.aps.org/doi/10.1103/PhysRevLett.48.291}.

\bibitem[{\citenamefont{Mermin}(1990)}]{Mermin1990}
\bibinfo{author}{\bibfnamefont{N.~D.} \bibnamefont{Mermin}},
  \bibinfo{journal}{Phys. Rev. Lett.} \textbf{\bibinfo{volume}{65}},
  \bibinfo{pages}{1838} (\bibinfo{year}{1990}),
  \urlprefix\url{https://link.aps.org/doi/10.1103/PhysRevLett.65.1838}.

\bibitem[{\citenamefont{G\"uhne et~al.}(2005)\citenamefont{G\"uhne, T\'oth,
  Hyllus, and Briegel}}]{Guehneetal2005}
\bibinfo{author}{\bibfnamefont{O.}~\bibnamefont{G\"uhne}},
  \bibinfo{author}{\bibfnamefont{G.}~\bibnamefont{T\'oth}},
  \bibinfo{author}{\bibfnamefont{P.}~\bibnamefont{Hyllus}}, \bibnamefont{and}
  \bibinfo{author}{\bibfnamefont{H.~J.} \bibnamefont{Briegel}},
  \bibinfo{journal}{Phys. Rev. Lett.} \textbf{\bibinfo{volume}{95}},
  \bibinfo{pages}{120405} (\bibinfo{year}{2005}),
  \urlprefix\url{https://link.aps.org/doi/10.1103/PhysRevLett.95.120405}.

\bibitem[{\citenamefont{Tura et~al.}(2017)\citenamefont{Tura, De~las Cuevas,
  Augusiak, Lewenstein, Ac{\'i}n, and Cirac}}]{tura_energy_2017}
\bibinfo{author}{\bibfnamefont{J.}~\bibnamefont{Tura}},
  \bibinfo{author}{\bibfnamefont{G.}~\bibnamefont{De~las Cuevas}},
  \bibinfo{author}{\bibfnamefont{R.}~\bibnamefont{Augusiak}},
  \bibinfo{author}{\bibfnamefont{M.}~\bibnamefont{Lewenstein}},
  \bibinfo{author}{\bibfnamefont{A.}~\bibnamefont{Ac{\'i}n}}, \bibnamefont{and}
  \bibinfo{author}{\bibfnamefont{J.}~\bibnamefont{Cirac}},
  \bibinfo{journal}{Phys. Rev. X} \textbf{\bibinfo{volume}{7}},
  \bibinfo{pages}{021005} (\bibinfo{year}{2017}),
  \urlprefix\url{https://link.aps.org/doi/10.1103/PhysRevX.7.021005}.

\bibitem[{\citenamefont{Lanyon et~al.}(2014)\citenamefont{Lanyon, Zwerger,
  Jurcevic, Hempel, D\"ur, Briegel, Blatt, and Roos}}]{Lanyonetal2014}
\bibinfo{author}{\bibfnamefont{B.~P.} \bibnamefont{Lanyon}},
  \bibinfo{author}{\bibfnamefont{M.}~\bibnamefont{Zwerger}},
  \bibinfo{author}{\bibfnamefont{P.}~\bibnamefont{Jurcevic}},
  \bibinfo{author}{\bibfnamefont{C.}~\bibnamefont{Hempel}},
  \bibinfo{author}{\bibfnamefont{W.}~\bibnamefont{D\"ur}},
  \bibinfo{author}{\bibfnamefont{H.~J.} \bibnamefont{Briegel}},
  \bibinfo{author}{\bibfnamefont{R.}~\bibnamefont{Blatt}}, \bibnamefont{and}
  \bibinfo{author}{\bibfnamefont{C.~F.} \bibnamefont{Roos}},
  \bibinfo{journal}{Phys. Rev. Lett.} \textbf{\bibinfo{volume}{112}},
  \bibinfo{pages}{100403} (\bibinfo{year}{2014}),
  \urlprefix\url{https://link.aps.org/doi/10.1103/PhysRevLett.112.100403}.

\bibitem[{\citenamefont{Tura et~al.}(2014)\citenamefont{Tura, Augusiak, Sainz,
  V\'ertesi, Lewenstein, and Ac\'i­n}}]{Turaetal2014}
\bibinfo{author}{\bibfnamefont{J.}~\bibnamefont{Tura}},
  \bibinfo{author}{\bibfnamefont{R.}~\bibnamefont{Augusiak}},
  \bibinfo{author}{\bibfnamefont{A.~B.} \bibnamefont{Sainz}},
  \bibinfo{author}{\bibfnamefont{T.}~\bibnamefont{V\'ertesi}},
  \bibinfo{author}{\bibfnamefont{M.}~\bibnamefont{Lewenstein}},
  \bibnamefont{and} \bibinfo{author}{\bibfnamefont{A.}~\bibnamefont{Ac\'i­n}},
  \bibinfo{journal}{Science} \textbf{\bibinfo{volume}{344}},
  \bibinfo{pages}{1256} (\bibinfo{year}{2014}), ISSN \bibinfo{issn}{0036-8075},
  \urlprefix\url{http://science.sciencemag.org/content/344/6189/1256}.

\bibitem[{\citenamefont{Fadel and Tura}(2017)}]{fadelT2017}
\bibinfo{author}{\bibfnamefont{M.}~\bibnamefont{Fadel}} \bibnamefont{and}
  \bibinfo{author}{\bibfnamefont{J.}~\bibnamefont{Tura}},
  \bibinfo{journal}{Phys. Rev. Lett.} \textbf{\bibinfo{volume}{119}},
  \bibinfo{pages}{230402} (\bibinfo{year}{2017}),
  \urlprefix\url{https://link.aps.org/doi/10.1103/PhysRevLett.119.230402}.

\bibitem[{\citenamefont{Wang et~al.}(2017)\citenamefont{Wang, Singh, and
  Navascu\'es}}]{wangN2017}
\bibinfo{author}{\bibfnamefont{Z.}~\bibnamefont{Wang}},
  \bibinfo{author}{\bibfnamefont{S.}~\bibnamefont{Singh}}, \bibnamefont{and}
  \bibinfo{author}{\bibfnamefont{M.}~\bibnamefont{Navascu\'es}},
  \bibinfo{journal}{Phys. Rev. Lett.} \textbf{\bibinfo{volume}{118}},
  \bibinfo{pages}{230401} (\bibinfo{year}{2017}),
  \urlprefix\url{https://link.aps.org/doi/10.1103/PhysRevLett.118.230401}.

\bibitem[{\citenamefont{Baccari et~al.}(2017)\citenamefont{Baccari, Cavalcanti,
  Wittek, and Ac\'{\i}n}}]{Baccarietal2017}
\bibinfo{author}{\bibfnamefont{F.}~\bibnamefont{Baccari}},
  \bibinfo{author}{\bibfnamefont{D.}~\bibnamefont{Cavalcanti}},
  \bibinfo{author}{\bibfnamefont{P.}~\bibnamefont{Wittek}}, \bibnamefont{and}
  \bibinfo{author}{\bibfnamefont{A.}~\bibnamefont{Ac\'{\i}n}},
  \bibinfo{journal}{Phys. Rev. X} \textbf{\bibinfo{volume}{7}},
  \bibinfo{pages}{021042} (\bibinfo{year}{2017}),
  \urlprefix\url{https://link.aps.org/doi/10.1103/PhysRevX.7.021042}.

\bibitem[{\citenamefont{Jaynes}(1957)}]{Jaynes1957}
\bibinfo{author}{\bibfnamefont{E.~T.} \bibnamefont{Jaynes}},
  \bibinfo{journal}{Phys. Rev.} \textbf{\bibinfo{volume}{106}},
  \bibinfo{pages}{620} (\bibinfo{year}{1957}),
  \urlprefix\url{https://link.aps.org/doi/10.1103/PhysRev.106.620}.

\bibitem[{\citenamefont{Tikochinsky et~al.}(1984)\citenamefont{Tikochinsky,
  Tishby, and Levine}}]{Tikochinskyetal1984}
\bibinfo{author}{\bibfnamefont{Y.}~\bibnamefont{Tikochinsky}},
  \bibinfo{author}{\bibfnamefont{N.~Z.} \bibnamefont{Tishby}},
  \bibnamefont{and} \bibinfo{author}{\bibfnamefont{R.~D.}
  \bibnamefont{Levine}}, \bibinfo{journal}{Phys. Rev. A}
  \textbf{\bibinfo{volume}{30}}, \bibinfo{pages}{2638} (\bibinfo{year}{1984}),
  \urlprefix\url{https://link.aps.org/doi/10.1103/PhysRevA.30.2638}.

\bibitem[{\citenamefont{Nguyen et~al.}(2017)\citenamefont{Nguyen, Zecchina, and
  Berg}}]{Nguyenetal2017}
\bibinfo{author}{\bibfnamefont{H.~C.} \bibnamefont{Nguyen}},
  \bibinfo{author}{\bibfnamefont{R.}~\bibnamefont{Zecchina}}, \bibnamefont{and}
  \bibinfo{author}{\bibfnamefont{J.}~\bibnamefont{Berg}},
  \bibinfo{journal}{Advances in Physics} \textbf{\bibinfo{volume}{66}},
  \bibinfo{pages}{197} (\bibinfo{year}{2017}),
  \eprint{https://doi.org/10.1080/00018732.2017.1341604},
  \urlprefix\url{https://doi.org/10.1080/00018732.2017.1341604}.

\bibitem[{\citenamefont{Boyd and Vandenberghe}(2004)}]{Boyd2004convex}
\bibinfo{author}{\bibfnamefont{S.}~\bibnamefont{Boyd}} \bibnamefont{and}
  \bibinfo{author}{\bibfnamefont{L.}~\bibnamefont{Vandenberghe}},
  \emph{\bibinfo{title}{Convex Optimization}} (\bibinfo{publisher}{Cambridge
  University Press}, \bibinfo{year}{2004}).

\bibitem[{\citenamefont{Eggert et~al.}(1994)\citenamefont{Eggert, Affleck, and
  Takahashi}}]{Eggertetal1994}
\bibinfo{author}{\bibfnamefont{S.}~\bibnamefont{Eggert}},
  \bibinfo{author}{\bibfnamefont{I.}~\bibnamefont{Affleck}}, \bibnamefont{and}
  \bibinfo{author}{\bibfnamefont{M.}~\bibnamefont{Takahashi}},
  \bibinfo{journal}{Phys. Rev. Lett.} \textbf{\bibinfo{volume}{73}},
  \bibinfo{pages}{332} (\bibinfo{year}{1994}),
  \urlprefix\url{https://link.aps.org/doi/10.1103/PhysRevLett.73.332}.

\bibitem[{\citenamefont{T\'oth}(2004)}]{Toth2004}
\bibinfo{author}{\bibfnamefont{G.}~\bibnamefont{T\'oth}},
  \bibinfo{journal}{Phys. Rev. A} \textbf{\bibinfo{volume}{69}},
  \bibinfo{pages}{052327} (\bibinfo{year}{2004}),
  \urlprefix\url{https://link.aps.org/doi/10.1103/PhysRevA.69.052327}.

\bibitem[{\citenamefont{Wie{\'{s}}niak
  et~al.}(2005)\citenamefont{Wie{\'{s}}niak, Vedral, and
  Brukner}}]{Wiesniaketal2005}
\bibinfo{author}{\bibfnamefont{M.}~\bibnamefont{Wie{\'{s}}niak}},
  \bibinfo{author}{\bibfnamefont{V.}~\bibnamefont{Vedral}}, \bibnamefont{and}
  \bibinfo{author}{\bibfnamefont{{\v{C}}.}~\bibnamefont{Brukner}},
  \bibinfo{journal}{New Journal of Physics} \textbf{\bibinfo{volume}{7}},
  \bibinfo{pages}{258} (\bibinfo{year}{2005}),
  \urlprefix\url{https://doi.org/10.1088%2F1367-2630%2F7%2F1%2F258}.

\bibitem[{SM()}]{SM}
\emph{\bibinfo{title}{{See Supplemental Material (SM) for a discussion of: 1)
  the implementation of the variational approach; 2) the reconstruction of the
  Bell inequality violated by the quantum critical point of the 2$d$ quantum
  Ising model; 3) the proof of the many-body Pearle-Braunstein-Caves
  inequality; 4) a detailed discussion of how quantum data coming from
  Heisenberg antiferromagnets can violate the above inequality and 5) the proof
  that this violation is the maximal one admitted by quantum mechanics. The SM
  contains the Refs.}~\cite{BloteD2002,Barahona1982,
  Bachas1984,Lucas2014,BinderLandau,MChandbook,Tauberbook,MoritaN2008,Katzgraberetal2001,Wangetal2014,Preskill2018}.}}

\bibitem[{\citenamefont{Clauser et~al.}(1969)\citenamefont{Clauser, Horne,
  Shimony, and Holt}}]{Clauseretal1969}
\bibinfo{author}{\bibfnamefont{J.~F.} \bibnamefont{Clauser}},
  \bibinfo{author}{\bibfnamefont{M.~A.} \bibnamefont{Horne}},
  \bibinfo{author}{\bibfnamefont{A.}~\bibnamefont{Shimony}}, \bibnamefont{and}
  \bibinfo{author}{\bibfnamefont{R.~A.} \bibnamefont{Holt}},
  \bibinfo{journal}{Phys. Rev. Lett.} \textbf{\bibinfo{volume}{23}},
  \bibinfo{pages}{880} (\bibinfo{year}{1969}),
  \urlprefix\url{https://link.aps.org/doi/10.1103/PhysRevLett.23.880}.

\bibitem[{\citenamefont{Schmied et~al.}(2016)\citenamefont{Schmied, Bancal,
  Allard, Fadel, Scarani, Treutlein, and Sangouard}}]{Schmiedetal2016}
\bibinfo{author}{\bibfnamefont{R.}~\bibnamefont{Schmied}},
  \bibinfo{author}{\bibfnamefont{J.-D.} \bibnamefont{Bancal}},
  \bibinfo{author}{\bibfnamefont{B.}~\bibnamefont{Allard}},
  \bibinfo{author}{\bibfnamefont{M.}~\bibnamefont{Fadel}},
  \bibinfo{author}{\bibfnamefont{V.}~\bibnamefont{Scarani}},
  \bibinfo{author}{\bibfnamefont{P.}~\bibnamefont{Treutlein}},
  \bibnamefont{and}
  \bibinfo{author}{\bibfnamefont{N.}~\bibnamefont{Sangouard}},
  \bibinfo{journal}{Science} \textbf{\bibinfo{volume}{352}},
  \bibinfo{pages}{441} (\bibinfo{year}{2016}), ISSN \bibinfo{issn}{0036-8075,
  1095-9203},
  \urlprefix\url{http://science.sciencemag.org/content/352/6284/441}.

\bibitem[{\citenamefont{Engelsen et~al.}(2017)\citenamefont{Engelsen,
  Krishnakumar, Hosten, and Kasevich}}]{Engelsenetal2017}
\bibinfo{author}{\bibfnamefont{N.~J.} \bibnamefont{Engelsen}},
  \bibinfo{author}{\bibfnamefont{R.}~\bibnamefont{Krishnakumar}},
  \bibinfo{author}{\bibfnamefont{O.}~\bibnamefont{Hosten}}, \bibnamefont{and}
  \bibinfo{author}{\bibfnamefont{M.~A.} \bibnamefont{Kasevich}},
  \bibinfo{journal}{Phys. Rev. Lett.} \textbf{\bibinfo{volume}{118}},
  \bibinfo{pages}{140401} (\bibinfo{year}{2017}),
  \urlprefix\url{https://link.aps.org/doi/10.1103/PhysRevLett.118.140401}.

\bibitem[{\citenamefont{Piga et~al.}(2019)\citenamefont{Piga, Aloy, Lewenstein,
  and Fr{\'e}rot}}]{Pigaetal2019}
\bibinfo{author}{\bibfnamefont{A.}~\bibnamefont{Piga}},
  \bibinfo{author}{\bibfnamefont{A.}~\bibnamefont{Aloy}},
  \bibinfo{author}{\bibfnamefont{M.}~\bibnamefont{Lewenstein}},
  \bibnamefont{and}
  \bibinfo{author}{\bibfnamefont{I.}~\bibnamefont{Fr{\'e}rot}},
  \bibinfo{journal}{Phys. Rev. Lett.} \textbf{\bibinfo{volume}{123}},
  \bibinfo{pages}{170604} (\bibinfo{year}{2019}),
  \urlprefix\url{https://link.aps.org/doi/10.1103/PhysRevLett.123.170604}.

\bibitem[{\citenamefont{Fr\'erot and Roscilde}(2018)}]{FrerotR2018}
\bibinfo{author}{\bibfnamefont{I.}~\bibnamefont{Fr\'erot}} \bibnamefont{and}
  \bibinfo{author}{\bibfnamefont{T.}~\bibnamefont{Roscilde}},
  \bibinfo{journal}{Phys. Rev. Lett.} \textbf{\bibinfo{volume}{121}},
  \bibinfo{pages}{020402} (\bibinfo{year}{2018}),
  \urlprefix\url{https://link.aps.org/doi/10.1103/PhysRevLett.121.020402}.

\bibitem[{\citenamefont{Pearle}(1970)}]{Pearle1970}
\bibinfo{author}{\bibfnamefont{P.~M.} \bibnamefont{Pearle}},
  \bibinfo{journal}{Phys. Rev. D} \textbf{\bibinfo{volume}{2}},
  \bibinfo{pages}{1418} (\bibinfo{year}{1970}),
  \urlprefix\url{https://link.aps.org/doi/10.1103/PhysRevD.2.1418}.

\bibitem[{\citenamefont{Braunstein and Caves}(1990)}]{BraunsteinC1990}
\bibinfo{author}{\bibfnamefont{S.~L.} \bibnamefont{Braunstein}}
  \bibnamefont{and} \bibinfo{author}{\bibfnamefont{C.~M.} \bibnamefont{Caves}},
  \bibinfo{journal}{Annals of Physics} \textbf{\bibinfo{volume}{202}},
  \bibinfo{pages}{22 } (\bibinfo{year}{1990}), ISSN \bibinfo{issn}{0003-4916},
  \urlprefix\url{http://www.sciencedirect.com/science/article/pii/000349169090339P}.

\bibitem[{\citenamefont{Georgescu et~al.}(2014)\citenamefont{Georgescu, Ashhab,
  and Nori}}]{Georgescuetal2013}
\bibinfo{author}{\bibfnamefont{I.~M.} \bibnamefont{Georgescu}},
  \bibinfo{author}{\bibfnamefont{S.}~\bibnamefont{Ashhab}}, \bibnamefont{and}
  \bibinfo{author}{\bibfnamefont{F.}~\bibnamefont{Nori}},
  \bibinfo{journal}{Rev. Mod. Phys.} \textbf{\bibinfo{volume}{86}},
  \bibinfo{pages}{153} (\bibinfo{year}{2014}),
  \urlprefix\url{https://link.aps.org/doi/10.1103/RevModPhys.86.153}.

\bibitem[{\citenamefont{{Altman} et~al.}(2019)\citenamefont{{Altman}, {Brown},
  {Carleo}, {Carr}, {Demler}, {Chin}, {DeMarco}, {Economou}, {Eriksson}, {Fu}
  et~al.}}]{Altmanetal2019}
\bibinfo{author}{\bibfnamefont{E.}~\bibnamefont{{Altman}}},
  \bibinfo{author}{\bibfnamefont{K.~R.} \bibnamefont{{Brown}}},
  \bibinfo{author}{\bibfnamefont{G.}~\bibnamefont{{Carleo}}},
  \bibinfo{author}{\bibfnamefont{L.~D.} \bibnamefont{{Carr}}},
  \bibinfo{author}{\bibfnamefont{E.}~\bibnamefont{{Demler}}},
  \bibinfo{author}{\bibfnamefont{C.}~\bibnamefont{{Chin}}},
  \bibinfo{author}{\bibfnamefont{B.}~\bibnamefont{{DeMarco}}},
  \bibinfo{author}{\bibfnamefont{S.~E.} \bibnamefont{{Economou}}},
  \bibinfo{author}{\bibfnamefont{M.~A.} \bibnamefont{{Eriksson}}},
  \bibinfo{author}{\bibfnamefont{K.-M.~C.} \bibnamefont{{Fu}}},
  \bibnamefont{et~al.}, \bibinfo{journal}{arXiv e-prints}
  \bibinfo{eid}{arXiv:1912.06938} (\bibinfo{year}{2019}), \eprint{1912.06938}.

\bibitem[{\citenamefont{Wendin}(2017)}]{Wendin2017}
\bibinfo{author}{\bibfnamefont{G.}~\bibnamefont{Wendin}},
  \bibinfo{journal}{Reports on Progress in Physics}
  \textbf{\bibinfo{volume}{80}}, \bibinfo{pages}{106001}
  (\bibinfo{year}{2017}),
  \urlprefix\url{https://doi.org/10.1088%2F1361-6633%2Faa7e1a}.

\bibitem[{\citenamefont{Bruzewicz et~al.}(2019)\citenamefont{Bruzewicz,
  Chiaverini, McConnell, and Sage}}]{Bruzewiczetal2019}
\bibinfo{author}{\bibfnamefont{C.~D.} \bibnamefont{Bruzewicz}},
  \bibinfo{author}{\bibfnamefont{J.}~\bibnamefont{Chiaverini}},
  \bibinfo{author}{\bibfnamefont{R.}~\bibnamefont{McConnell}},
  \bibnamefont{and} \bibinfo{author}{\bibfnamefont{J.~M.} \bibnamefont{Sage}},
  \bibinfo{journal}{Applied Physics Reviews} \textbf{\bibinfo{volume}{6}},
  \bibinfo{pages}{021314} (\bibinfo{year}{2019}),
  \eprint{https://doi.org/10.1063/1.5088164},
  \urlprefix\url{https://doi.org/10.1063/1.5088164}.

\bibitem[{\citenamefont{Bl\"ote and Deng}(2002)}]{BloteD2002}
\bibinfo{author}{\bibfnamefont{H.~W.~J.} \bibnamefont{Bl\"ote}}
  \bibnamefont{and} \bibinfo{author}{\bibfnamefont{Y.}~\bibnamefont{Deng}},
  \bibinfo{journal}{Phys. Rev. E} \textbf{\bibinfo{volume}{66}},
  \bibinfo{pages}{066110} (\bibinfo{year}{2002}),
  \urlprefix\url{https://link.aps.org/doi/10.1103/PhysRevE.66.066110}.

\bibitem[{\citenamefont{Barahona}(1982)}]{Barahona1982}
\bibinfo{author}{\bibfnamefont{F.}~\bibnamefont{Barahona}},
  \bibinfo{journal}{Journal of Physics A: Mathematical and General}
  \textbf{\bibinfo{volume}{15}}, \bibinfo{pages}{3241} (\bibinfo{year}{1982}),
  \urlprefix\url{https://doi.org/10.1088%2F0305-4470%2F15%2F10%2F028}.

\bibitem[{\citenamefont{Bachas}(1984)}]{Bachas1984}
\bibinfo{author}{\bibfnamefont{C.~P.} \bibnamefont{Bachas}},
  \bibinfo{journal}{Journal of Physics A: Mathematical and General}
  \textbf{\bibinfo{volume}{17}}, \bibinfo{pages}{L709} (\bibinfo{year}{1984}),
  \urlprefix\url{https://doi.org/10.1088%2F0305-4470%2F17%2F13%2F006}.

\bibitem[{\citenamefont{Lucas}(2014)}]{Lucas2014}
\bibinfo{author}{\bibfnamefont{A.}~\bibnamefont{Lucas}},
  \bibinfo{journal}{Frontiers in Physics} \textbf{\bibinfo{volume}{2}},
  \bibinfo{pages}{5} (\bibinfo{year}{2014}), ISSN \bibinfo{issn}{2296-424X},
  \urlprefix\url{https://www.frontiersin.org/article/10.3389/fphy.2014.00005}.

\bibitem[{\citenamefont{Landau and Binder}(2014)}]{BinderLandau}
\bibinfo{author}{\bibfnamefont{D.~P.} \bibnamefont{Landau}} \bibnamefont{and}
  \bibinfo{author}{\bibfnamefont{K.}~\bibnamefont{Binder}},
  \emph{\bibinfo{title}{A Guide to Monte Carlo Simulations in Statistical
  Physics}} (\bibinfo{publisher}{Cambridge University Press},
  \bibinfo{year}{2014}).

\bibitem[{\citenamefont{Brooks et~al.}(2011)\citenamefont{Brooks, Gelman,
  Jones, and Meng}}]{MChandbook}
\bibinfo{author}{\bibfnamefont{S.}~\bibnamefont{Brooks}},
  \bibinfo{author}{\bibfnamefont{A.}~\bibnamefont{Gelman}},
  \bibinfo{author}{\bibfnamefont{G.~L.} \bibnamefont{Jones}}, \bibnamefont{and}
  \bibinfo{author}{\bibfnamefont{X.-L.~E.} \bibnamefont{Meng}},
  \emph{\bibinfo{title}{Handbook of Markov Chain Monte Carlo}}
  (\bibinfo{publisher}{CRC Press, Boca Raton}, \bibinfo{year}{2011}).

\bibitem[{\citenamefont{Tauber}(2014)}]{Tauberbook}
\bibinfo{author}{\bibfnamefont{U.~C.} \bibnamefont{Tauber}},
  \emph{\bibinfo{title}{Critical Dynamics}} (\bibinfo{publisher}{Cambridge
  University Press}, \bibinfo{year}{2014}).

\bibitem[{\citenamefont{Morita and Nishimori}(2008)}]{MoritaN2008}
\bibinfo{author}{\bibfnamefont{S.}~\bibnamefont{Morita}} \bibnamefont{and}
  \bibinfo{author}{\bibfnamefont{H.}~\bibnamefont{Nishimori}},
  \bibinfo{journal}{Journal of Mathematical Physics}
  \textbf{\bibinfo{volume}{49}}, \bibinfo{pages}{125210}
  (\bibinfo{year}{2008}), \eprint{https://doi.org/10.1063/1.2995837},
  \urlprefix\url{https://doi.org/10.1063/1.2995837}.

\bibitem[{\citenamefont{Katzgraber et~al.}(2001)\citenamefont{Katzgraber,
  Palassini, and Young}}]{Katzgraberetal2001}
\bibinfo{author}{\bibfnamefont{H.~G.} \bibnamefont{Katzgraber}},
  \bibinfo{author}{\bibfnamefont{M.}~\bibnamefont{Palassini}},
  \bibnamefont{and} \bibinfo{author}{\bibfnamefont{A.~P.} \bibnamefont{Young}},
  \bibinfo{journal}{Phys. Rev. B} \textbf{\bibinfo{volume}{63}},
  \bibinfo{pages}{184422} (\bibinfo{year}{2001}),
  \urlprefix\url{https://link.aps.org/doi/10.1103/PhysRevB.63.184422}.

\bibitem[{\citenamefont{Wang et~al.}(2015)\citenamefont{Wang, Machta, and
  Katzgraber}}]{Wangetal2014}
\bibinfo{author}{\bibfnamefont{W.}~\bibnamefont{Wang}},
  \bibinfo{author}{\bibfnamefont{J.}~\bibnamefont{Machta}}, \bibnamefont{and}
  \bibinfo{author}{\bibfnamefont{H.~G.} \bibnamefont{Katzgraber}},
  \bibinfo{journal}{Phys. Rev. E} \textbf{\bibinfo{volume}{92}},
  \bibinfo{pages}{013303} (\bibinfo{year}{2015}),
  \urlprefix\url{https://link.aps.org/doi/10.1103/PhysRevE.92.013303}.

\bibitem[{\citenamefont{Preskill}(2018)}]{Preskill2018}
\bibinfo{author}{\bibfnamefont{J.}~\bibnamefont{Preskill}},
  \bibinfo{journal}{{Quantum}} \textbf{\bibinfo{volume}{2}},
  \bibinfo{pages}{79} (\bibinfo{year}{2018}), ISSN \bibinfo{issn}{2521-327X},
  \urlprefix\url{https://doi.org/10.22331/q-2018-08-06-79}.

\end{thebibliography}

\null

\newpage

\begin{center}
{\bf SUPPLEMENTAL MATERIAL} \\
{\bf Detecting many-body Bell nonlocality by solving Ising models} \\

\vspace{.5cm}
Ir\'en\'ee Fr\'erot$^{1,2}$ and Tommaso Roscilde$^{3}$ 

\vspace{.5cm}
$^1$ ICFO-Institut de Ciencies Fotoniques, The Barcelona Institute of Science and Technology, Av. Carl Friedrich Gauss 3, 08860 Barcelona, Spain \\
$^2$ Max-Planck-Institut f{\"u}r Quantenoptik, D-85748 Garching, Germany \\
$^3$ Laboratoire de Physique, CNRS UMR 5672, Ecole Normale Sup\'erieure de Lyon, Universit\'e de Lyon, 46 All\'ee d'Italie, Lyon, F-69364, France 

\end{center}

\begin{center}
In this Supplemental Material, we detail the Monte-Carlo implementation of the variational search of Local-Variable models, and discuss its scalability. We then expose in details its application to quantum data obtained from measurements on the $2d$ quantum Ising model at its quantum-critical point. Finally, we discuss the many-body Pearle-Braustein-Caves Bell inequality that we discovered running our algorithm on the correlation functions in the ground state of Heisenberg antiferromagnets.
\end{center}

\renewcommand{\theequation}{S\arabic{equation}}
\renewcommand{\thefigure}{S\arabic{figure}}

\section{Implementation of the variational approach}

 In this section we describe the practical implementation of the algorithm for the search of Bell inequalities as starting from a set of thoughtfully chosen quantum data. More elements about the choice of the quantum data -- specifically for what concerns the choice of the measurement basis -- will be discussed in Secs. ~\ref{e.Ising} and  \ref{s.PBCviolation}.

  \subsection{Monte Carlo treatment of the local-variable theory}
  
  The goal of the algorithm is to reproduce the quantum data using the equilibrium behavior of a classical statistical physics model describing Ising variables (for $p=2$) on a lattice, or more generally with variables admitting an arbitrary number $p$ of values. As discussed in the main text, this represent an efficient strategy to produce a local-variable (LV) theory for the data of interest. In the following we will concentrate for concreteness on the $(N,k,p)$ scenario with $p=2$, but the discussion can be readily generalized to arbitrary $p$'s.  The quantum data generically consists of average values of functions $f_r(\bm \sigma)$ of the measurement outputs $\{\sigma_a^{(i)}\}$ ($a = 0, ..., k-1$, $i = 1,...N$) on the quantum degrees of freedom, which we shall denote as qubits in the following (since $p=2$); such functions are then used to build the Hamiltonian of the classical model in the form:
  \begin{equation}
  {\cal H}(\bm \sigma;\bm K)= -\sum_r K_r  f_r(\bm \sigma)~.
  \label{e.Heff}
  \end{equation}
  Throughout the work presented here, we have specialized our attention to single-site expectation values and two-point correlation functions, namely $\{f_r(\bm \sigma)\} = \{ \sigma_a^{(i)} \}, \{ \sigma_a^{(i)} \sigma_b^{(j\neq i)} \}$; but obviously the whole treatment generalizes to arbitrary correlation functions. 
  
 The core of the variational algorithm is the calculation of the statistical averages representing the predictions of the LV theory
 \begin{equation}
 \langle f_r(\bm \sigma) \rangle_{\rm LV} = \frac{1}{{\cal Z}(\bm K)} \sum_{\bm \sigma} f_r(\bm \sigma) e^{-{\cal H}(\bm \sigma;\bm K)}
 \label{e.average_f}
 \end{equation}
and building up the gradient of the cost function (Eq.~5 of the main text)
\begin{equation}
G_r = \langle  f_r(\bm \sigma) \rangle_{\rm LV} -  \langle f_r(\bm \sigma) \rangle_{\rm Q} 
\label{e.grad}
\end{equation}
which is then used to update the coupling constants within a gradient-descent algorithm. In our calculations we have actually implemented Nesterov's accelerated gradient descent, which offers a dramatic speed-up, often reducing the cost function exponentially in the number of iterations -- yet with the mild drawback that the cost function is not necessarily decreasing at each step of the algorithm. 
  
  The effective Hamiltonian of Eq.~\eqref{e.Heff} has a structure involving generically long-range couplings and external fields (see below); a general, fully scalable strategy to calculate efficiently the statistical averages of Eq.~\eqref{e.average_f} is then to use a standard Monte Carlo (MC) algorithm with single spin-flip Metropolis updates. Defining a single MC step as ${\cal O}(N)$ spin-flip attempts, the statistical error  ${\rm Err}(\langle  f_r(\bm \sigma) \rangle_{\rm LV})$ on the LV averages  $\langle  f_r(\bm \sigma) \rangle_{\rm LV}$ scales as $ c ~N_{\rm MC}^{-1/2}$ where $N_{\rm MC}$ is the number of MC steps and $c$ is a size-independent prefactor.
  
   In a practical calculation aimed at optimizing the LV theory,  one in fact needs to have a good relative precision not on the LV observables as such, but rather on the gradient as in Eq.~\eqref{e.grad} -- for instance one may require that 
  \begin{equation}
  \frac{{\rm Err}(|\bm G|^2)}{|\bm G|^2} = \frac{\sum_{r=1}^R  2~{\rm Err}(\langle  f_r(\bm \sigma) \rangle_{\rm LV}) ~|G_r|}{\sum_r |G_r|^2} \leq \eta  
  \end{equation}	
for a given tolerance $\eta$, in order for the gradient to be calculated with sufficient precision so as to offer practical guidance in the optimization. For the time being we have neglected the uncertainty on the quantum data, assuming that it is much smaller than the statistical uncertainty on the LV averages. This means that the number of Monte Carlo steps scales in practice as 
$N_{\rm MC} \sim (\eta |\bm G|^2)^{-2}$ , namely the cost of the MC simulation grows as the distance $|\bm G|$ between the LV predictions and the quantum data decreases. 

The above result on the scaling of the MC steps with the gradient norm would naively imply that, if the LV theory succeeds to reproduce the quantum data ($|\bm G|\to 0$), then the cost of the computation of the LV averages would diverge. This is in fact not the case, when taking into account an inevitable finite precision on the quantum data, namely that ${\rm Err}(\langle  f_r(\bm \sigma) \rangle_{\rm Q}) > 0$. In that case, convergence of the LV theory occurs \emph{within} the precision of the quantum data, and this imposes a minimal error on the LV averages of the same order of magnitude. Hence the maximum number of MC steps necessary to assess convergence of the LV theory to the quantum data is of the order $N_{\rm MC} \sim \max_r [{\rm Err}(\langle  f_r(\bm \sigma) \rangle_{\rm Q})^{-2}]$.

Otherwise, if the LV theory fails to reproduce the quantum data, the gradient norm saturates to a finite minimum $|{\bm G}_{\infty}|$; and the number of MC steps necessary to reach this conclusion saturates to $N_{\rm MC} \sim (\eta |\bm G|^2_{\infty})^{-2}$. In our calculations we have found that a tolerance $\eta =0.05$ provides a manageable cost (with $N_{\rm MC}\sim 10^6\div 10^7$ for the systems we considered) while providing a sufficient accuracy on the gradient for the optimization to be effective. 
	
    \subsection{Effective local-variable Hamiltonian structured by the quantum data}
    
  The structure of the quantum data $\langle f_r(\bm \sigma) \rangle_{\rm Q}$ that we aim at reproducing fundamentally consists of two ingredients: 1) the geometry of the observables collected on the qubits (namely, what correlation functions among which qubits have been measured, etc.) ; and 2) the spatial structure of the measured observables (namely, what values the correlation functions actually take). Such properties of the quantum data fundamentally dictate the structure and physical properties of the effective Hamiltonian to be treated. In the following we shall make some conservative assumptions on both aspects of the structure of quantum data, which guarantee the success of our approach; and we shall briefly discuss the possible lifting of such assumptions. To avoid any confusion, let us stress once again that the inverse problems one has to solve in our approach are purely \textit{classical} ones. The following discussion is dedicated to the computational resources required to solve them as problems in classical statistical physics at equilibrium.
  
 \emph{Computational hardness of Ising spin models.}  
The central computational task of our variational search for LV theories is the calculation of thermodynamic expectation values for generalized Ising models.
By construction, the Monte Carlo method builds a Markov chain which samples a Boltzmann distribution so as to predict statistical average values with  a precision scaling with the inverse of the square root of the number of steps \cite{BinderLandau, MChandbook}. The time of convergence of the Markov chain to its stationary regime is not predicted in general, but it is typically observed to be at most polynomial in the system size (as the result of the number of elementary updates necessary to de-correlate two successive configurations). If one Monte Carlo time step involves an extensive number of elementary updates ${\cal O}(N)$, the typical number of step necessary to relax the system scales at most as $N^{z/d}$ in a $d$-dimensional system, where $z$ is the so-called dynamical critical exponent, which is zero away from transitions points and it is finite (and typically ${\cal O}(1)$ \cite{Tauberbook}) at second-order transitions. Hence at most a global scaling of the convergence time as $O(N^{1+z/d})$ is expected.  

  As such the Monte Carlo method could be considered \emph{a priori} efficient in reconstructing the equilibrium statistical physics of any classical model in polynomial time. This conclusion, nonetheless, is at odds with the proof that the ground-state search for some classes of Ising \emph{spin glasses}, namely models with random and frustrated interactions, is in fact a NP-hard problem \cite{Barahona1982, Bachas1984}. Moreover many hard combinatorial problems (in the NP-hard or even NP-complete class) can be cast in the form of the ground-state search of an Ising model \cite{Lucas2014}; hard instances of such combinatorial problems also realize examples of Ising spin glasses.  This suggests that an efficient solution (namely arbitrarily scalable in polynomial time) of their low-temperature thermodynamics using Monte Carlo is highly unlikely, as it would amount to proving that P$\equiv$NP. In practice, the calculation of the equilibrium expectation values for such models at low temperature is faced with the existence of a large number of metastable states, so that the convergence of MC simulations towards the correct Boltzmann averages (Eq.~\eqref{e.average_f}) might only be guaranteed in a time scaling exponentially with $N$ (using \emph{e.g.} simulated annealing \cite{MoritaN2008}).  We are not aware of other classes of Ising models (outside of Ising spin glasses) for which the ground-state search is proven to be NP-hard.
 
 In such a context, it is fair (and rather common) to adopt the working assumption that the Monte Carlo method provides an efficient calculation of the thermodynamics for all models which are \emph{not} proven to be NP-hard -- essentially for all models that are not spin glasses.  This assumption is corroborated \emph{e.g.} by decades of successful Monte Carlo investigations of Ising models on regular lattices.
Therefore we implicitly assume in the following that for all models which do \emph{not} feature simultaneously frustration and randomness (the two crucial ingredients of spin glasses) thermodynamic expectation values can be efficiently calculated using Monte Carlo, with the scaling properties which are predicted by the general theory of Monte Carlo methods \cite{BinderLandau, MChandbook}. The goal of the following paragraphs is to define minimal requirements on the quantum data that prevent the variational search of an LV description from potentially dealing with the thermodynamics of computationally hard Ising spin glasses.

  \emph{Geometry of the observables.} We shall assume some basic regularity in the choice of observables used to probe the quantum system -- namely the fact that, once the local measurement basis has been chosen (not necessarily in the same way for all qubits), then the same amount of information is acquired on each qubit and on its correlations to the other qubits. Arranging the qubits on a regular lattice with site index $i$, we shall imagine for instance that  the quantum data contain the average value of all the measurements on individual qubits $\langle \sigma_a^{(i)} \rangle_{\rm Q}$ for \emph{all} $a$ and $i$; and all the correlation functions among local measurements   
 on qubits lying within a given mutual distance $D$, namely $\langle \sigma_a^{(i)} \sigma_b^{(j)} \rangle_{\rm Q}$ for \emph{all} $a$, $b$ and for \emph{all} $i\neq j$ such that $|i-j| \leq D$, etc. Such a choice endows then the classical model with a regular lattice structure, with local field terms $- K_a^{(i)} \sigma_a^{(i)}$ coupling to each local variable, and with interaction terms $- K_{ab}^{(ij)} \sigma_a^{(i)} \sigma_b^{(j)}$ among all local variables within a given mutual distance. Non-locality is best detected when using as much information as possible on the quantum state; in particular, given access to the $n$-point correlation function, it is sensible to use the information on its entire structure, containing ${\cal O}(N^n)$ terms. Under this assumption, the LV effective Hamiltonian Eq.~\eqref{e.Heff} aimed at reproducing the full $n$-point correlation function contains a number ${\cal O}(N^n)$ of terms, implying a computational cost scaling in the same way with $N$ for the MC evaluation of the LV averages. In all the calculations presented in this work we have restricted our attention to the case $n=2$.

 \emph{Spatial structure of the measured observables.} Once the geometry of the coupling constants $\{ K_r \}$ has been fixed by the geometry of the observables contained in the quantum data, the values of such constants are adjusted in order to best reproduce the quantum data with an LV theory. Despite the regularity of the lattice into which the qubits are arranged, a preparation of their quantum state using random protocols (the equilibrium or non-equilibrium physics of random quantum Hamiltonians; random sequences of gates, etc.) will lead to quantum data with little or no structure, which may in turn require random couplings $K_r$  in the LV theory that aims at reproducing them. Avoiding randomness is a rather safe assumption in order to guarantee convergence for the Monte Carlo calculations of the LV averages (see below for further discussion). We can state that our approach is fully scalable -- namely capable of testing non-locality for the state of $N$ qubits in a time scaling polynomially with $N$ -- when considering quantum data endowed with basic symmetries (\emph{e.g.} translational invariance on a lattice with periodic boundary conditions, reflection symmetry on a regular lattice with open boundary conditions, etc.). Throughout this work we have used quantum data which come from the equilibrium physics of translationally invariant models (defined on regular lattices with periodic boundary conditions), which is a rather convenient situation as it allows one to reduce the number of independent coupling constants by a factor of $N$.  
 
 \emph{Concluding remarks.} Relaxing some of the above assumptions (on the regularity on the choice of observables building up the quantum data, and on the symmetries of the quantum data) may potentially expose the variational search of the LV theory to a much higher computational cost, which is required in the case in which the effective Hamiltonian  Eq.~\eqref{e.Heff} is a model of an Ising spin glass -- namely a model containing frustration \emph{and} randomness. While frustration is a generic ingredient of LV theories aiming at reproducing a set of quantum data, randomness is precisely avoided by the above assumptions. 
 Nonetheless it is very difficult to assess a priori the computational cost in the calculation of statistical averages for the frustrated and disordered Ising models that correspond to LV theories aiming  at reproducing a particular set of quantum data with randomness. We take the above assumptions on the structure of quantum data as being \emph{sufficient} (namely conservative) conditions for the convergence of our approach in a time scaling polynomially with system size. At the same time, one may reasonably expect that a moderate amount of randomness in the quantum data (either stemming from a random choice of the observables, or from randomness in the quantum state) can be tolerated without generating an exponential cost for the convergence of our approach. Even in the case of known intractable models, refined Monte Carlo approaches (such as simulated annealing, parallel tempering and population annealing) exist that allow one to accurately calculate the equilibrium properties on moderate-size lattices -- as it has been widely demonstrated in the case of Ising spin glasses \cite{Katzgraberetal2001, Wangetal2014}. These approches can be directly integrated within our variational optimization algorithm, opening the possibility of assessing non-locality in \emph{arbitrary} sets of quantum data (namely with arbitrarily random structure) coming from systems of moderate size. This perspective is highly appealing, as it provides the possibility of a systematic, device-independent certification of entanglement in most near-term quantum devices (so-called NISQ ones, for ``noisy intermediated-scale quantum" \cite{Preskill2018}).   
  
\section{Detecting non-locality of the 2$d$ transverse-field Ising model at the quantum critical point}
\label{e.Ising}

 In this section we shall focus on the quantum critical point of the 2$d$ quantum Ising model; and we shall describe how our algorithm recovers the many-body Bell inequality originally discovered in Ref.~\cite{Turaetal2014} as the Bell inequality which is most strongly violated by quantum data consisting of one-site and two-site expectation values in a $(N,2,2)$ scenario.  

  \begin{figure*}
\includegraphics[width=0.8\linewidth]{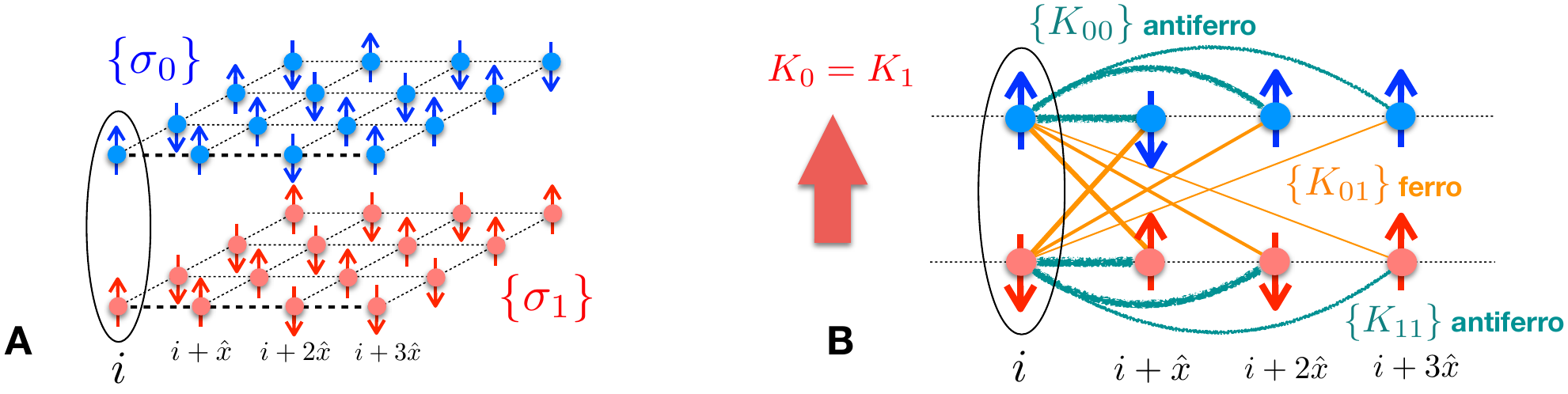}
\caption{Building the LV theory for the 2$d$ quantum Ising model at its quantum critical point. A:  in the $(N,2,2)$ scheme each lattice has attached two spin variables $\sigma_0^{(i)}$ and $\sigma_1^{(i)}$, so that the corresponding LV theory corresponds to an Ising model on a bilayer. B: sketch of the couplings emanating from a lattice $i$ along the $x$ direction of the lattice; in order to account for the fact that 
$\langle \sigma_0^{(i)} \sigma_1^{(j)} \rangle_{\rm Q} >  \langle \sigma_0^{(i)} \sigma_0^{(j)} \rangle_{\rm Q} = \langle \sigma_1^{(i)} \sigma_1^{(j)} \rangle_{\rm Q}$ in the presence of a strong polarization induced by the field $K_0 = K_1$, the LV theory develops frustration for the $K_{00}$  and $K_{11}$ couplings, as they are antiferromagnetic ($K^{(ij)}_{00} = K^{(ij)}_{11} < 0, ~\forall i\neq j $) and long-ranged, while the $K_{01}$ couplings are ferromagnetic and therefore unfrustrated  ($K^{(ij)}_{01}>  0, ~\forall i\neq j $) .}
\label{Fig1_SM}
\end{figure*}

\subsection{Frustration of the LV theories from squeezing and polarization}
 
  The Hamiltonian of the 2$d$ quantum Ising model reads
  \begin{equation}
  \hat{\cal H} = -J \sum_{\langle ij \rangle} \hat{S}_z^{(i)} \hat{S}_z^{(j)} - \Gamma \sum_i \hat{S}_x^{(i)}
  \end{equation} 
  where $\langle ij \rangle$ represents a pair of nearest neighbors on a $L\times L$ square lattice with periodic boundary conditions, and $\hat{S}_\alpha^{(i)}$ ($\alpha = x,y,z$) are $S=1/2$ spin operators. For definiteness, in the following we shall consider ferromagnetic couplings ($J>0$). The above model exhibits a quantum phase transition when the transverse field hits the critical value $\Gamma_c = 1.52219(1) J$ \cite{BloteD2002}: the critical point is accompanied with squeezing of the fluctuations of the $y$ component of the collective spin, $\hat J_y = \sum_i \hat S_y^{(i)}$ \cite{FrerotR2018}, namely with the property that the spin-spin correlation for the $y$ component of the spins is negative, $C_{yy}^{(ij)} = \langle \hat S_y^{(i)} \hat S_y^{(j)} \rangle < 0 ~~\forall i\neq j$, while the correlation functions  $C_{xx}$ and $C_{zz}$ for the $x$ and $z$ spin components are clearly positive. The appearance of sufficiently strong squeezing leads to Bell non-locality in a $(N,2,2)$ scenario, as shown experimentally in Ref.~\cite{Schmiedetal2016} for the squeezed state of an atomic spin ensemble; in the latter reference the following uniform measurement strategy was suggested
  \begin{equation}
  \hat \sigma_0 = \cos\theta ~\hat\sigma_x + \sin\theta ~\hat\sigma_y ~~~~~~~ \hat \sigma_1 = \cos\theta ~\hat\sigma_x - \sin\theta ~\hat\sigma_y
  \end{equation}
  where $\hat\sigma_{x,y}$ are Pauli matrices. Such a strategy was used to build a Bell non-locality witness -- namely an inequality on observables derived from the Bell inequality of Ref.~\cite{Turaetal2014} upon assuming the spin algebra -- based on the squeezing parameter $\xi^2 = N {\rm Var}(\hat J_y)/\langle \hat J_x\rangle^2$ and on the polarization $m_x = \langle \hat J_x\rangle/N = \langle \sum_i \hat S_x^{(i)} \rangle/N$. 
  
  As input to the Bell test we shall use the quantum data composed of the average local observables and of their two-point correlation functions, namely
 \begin{eqnarray}
 \langle \sigma_0^{(i)} \rangle_{\rm Q} & = & \langle \sigma_1^{(i)} \rangle_{\rm Q} = 2 \cos\theta~ m_x \nonumber \\
 \langle \sigma_0^{(i)} \sigma_0^{(j)} \rangle_{\rm Q} & = & \langle \sigma_1^{(i)} \sigma_1^{(j)} \rangle_{\rm Q} = 4 \cos^2\theta ~C_{xx}^{(ij)} + 4 \sin^2\theta~ C_{yy}^{(ij)} \nonumber \\
 \langle \sigma_0^{(i)} \sigma_1^{(j)}\rangle_{\rm Q} & = &  4 \cos^2\theta~ C_{xx}^{(ij)} - 4 \sin^2\theta ~C_{yy}^{(ij)}
 \label{e.quantumdata}
 \end{eqnarray}
  where we have used the fact that $\langle \hat S_x^{(i)} \hat S_y^{(j)} \rangle = 0$. From the point of view of our variational approach, the Bell test consists in actively building an LV theory in the form of an Ising model with two Ising variables ($\sigma_0^{(i)}$ and $\sigma_1^{(i)}$) per physical lattice site -- see Fig.~\ref{Fig1_SM}(A). In practice the LV theory aims at reproducing the quantum data via the equilibrium thermodynamics of a classical Ising model in an external field, with an effective Hamiltonian directly reflecting the content of the quantum data, namely containing all possible one-spin and two-spin terms
   \begin{eqnarray}
   {\cal H}(\bm \sigma;\bm K) & = & -\sum_i \left ( K_0 \sigma_0^{(i)} + K_1 \sigma_1^{(i)} \right ) \nonumber \\
    &&- \sum_{i<j} \Big [ K^{|i-j|}_{00} \sigma_0^{(i)} \sigma_0^{(j)} + K^{|i-j|}_{11} \sigma_1^{(i)} \sigma_1^{(j)} \nonumber \\
    && ~~~~~~~~ + K^{|i-j|}_{01} \left (\sigma_0^{(i)} \sigma_1^{(j)} + \sigma_1^{(i)} \sigma_0^{(j)} \right ) \Big ] ~.
    \label{e.Heff_Ising}
   \end{eqnarray}
  Here the translational invariance of the coupling constants (uniform fields $K_0$ and $K_1$, and $K_{ab}$ couplings uniquely dependent on the distance $|i-j|$ between sites) descends from the same invariance in the quantum data.  
  
  A simple argument suggests that the terms in the above effective Hamiltonian must be competing in energy (and therefore frustrated) in order to reproduce the quantum data of Eq.~\eqref{e.quantumdata}. Indeed the quantum data impose on the classical Ising variables in the Hamiltonian \eqref{e.Heff_Ising} seemingly contradictory requirements: the local variables should exhibit the same polarization  $\langle \sigma_0 \rangle =  \langle \sigma_1 \rangle$ and the same \emph{intra}-variable correlation functions $\langle \sigma_0^{(i)} \sigma_0^{(j)}\rangle = \langle \sigma_1^{(i)} \sigma_1^{(j)} \rangle$; yet, because of the negativity of the $C_{yy}$ correlator, the quantum data exhibit different \emph{inter}-variable correlation functions, namely $\langle \sigma_0^{(i)} \sigma_1^{(j)}\rangle > \langle \sigma_0^{(i)} \sigma_0^{(j)} \rangle > 0$. Such a situation could be in principle stabilized at finite temperature by simply taking different coupling constants $K_{01} > K_{00} = K_{11}$, so that thermal effects have different impact on the inter- vs. intra-variable correlations. Yet, if the polarization of the variables ($2 \cos\theta ~m_x$) is sufficiently strong, such thermal effects should be significantly suppressed by the fields $K_0$ and $K_1$, and the difference between intra- and inter-variable correlations is not justified. 
  
  The way out of this contradiction is that the coupling constants $K_{00}, K_{11}$ have opposite signs with respect to the couplings $K_{01}$; in particular, given the hierarchy of correlations,  the $K_{01}$ couplings should be taken as ferromagnetic (namely positive) so as to stabilize the positive dominant $\langle \sigma_0  \sigma_1 \rangle$ correlations; while the  $K_{00}$ and $K_{11}$ couplings should be taken as anti-ferromagnetic (negative), so as to suppress the $\langle \sigma_0 \sigma_0 \rangle$ and $\langle \sigma_1 \sigma_1 \rangle$ correlations. In particular the long-range nature of the latter antiferromagnetic couplings leads to frustration -- see Fig.~\ref{Fig1_SM}(B). Such a frustrated configuration of couplings is indeed the one corresponding to the unique global minimum of the cost function optimized in order for the LV theory to best reproduce the quantum data.

 \begin{figure}
\includegraphics[width=0.8\columnwidth]{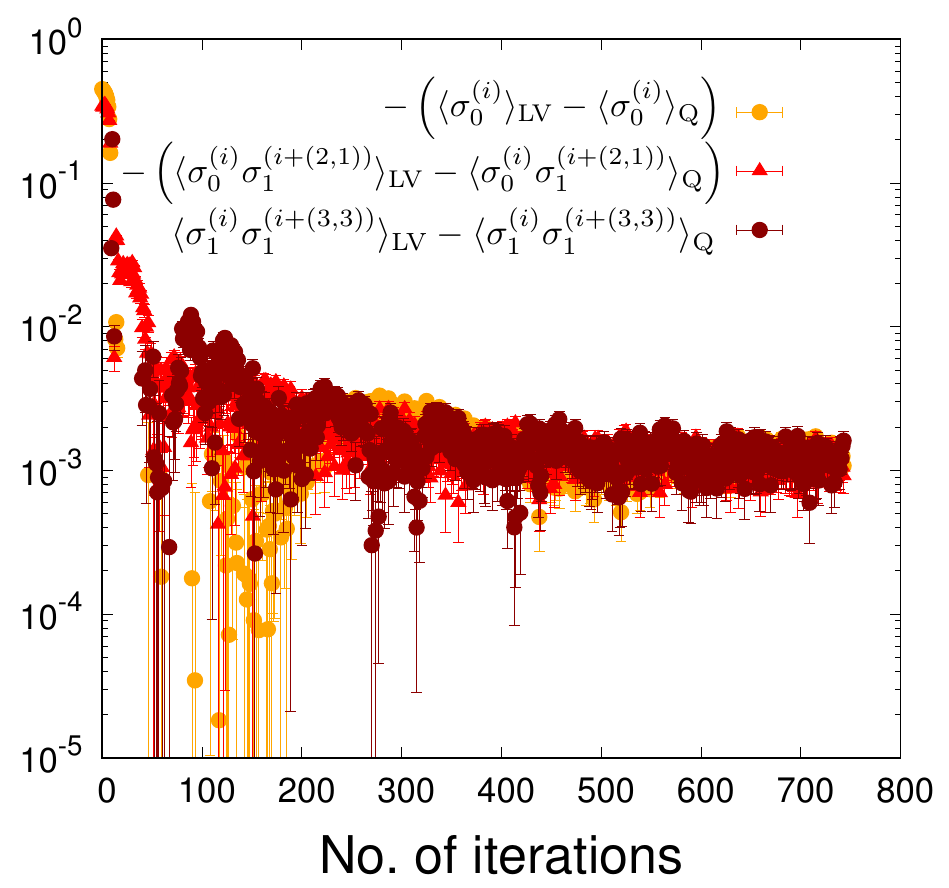}
\caption{Convergence of the gradient components of the cost function, $G_r = \langle f_r \rangle_{\rm LV} - \langle f_r \rangle_{\rm Q}$,  to their asymptotic value, corresponding to the minimum of the cost function.
The data shown correspond to three representative components for the distance to the quantum data of the 2d quantum Ising model on a 6x6 lattice at its quantum critical point, and for the angle $\theta = 0.3 \pi$.}
\label{Fig2_SM}
\end{figure}

\subsection{Convergence of the algorithm and emergence of the Bell inequality}
 
 Here we shall illustrate how the variational algorithm indicates the failure of the LV theory to reproduce the quantum data, and how it unveils the Bell inequality most strongly violated by the quantum data. 
 
 Fig.~\ref{Fig2_SM} shows the evolution of a few representative components of the cost function gradient (namely the distance vector between the LV predictions and the quantum data) as calculated for the 2$d$ quantum Ising model on a $6\times 6$ lattice at its quantum critical point. The evolution follows an accelerated gradient descent with increment $\epsilon = 10^{-2}$.  We observe that all the gradient components saturate to a finite asymptotic value, clearly manifesting a runaway behavior of the minimum search for the cost function (see main text). Moreover, and most significantly, all gradient components seemingly converge to the \emph{same} absolute value, with a specific sign pattern. Such a sign pattern reconstructs a Bell inequality of the following form 
 \begin{eqnarray}
  & - & \sum_i \left ( \sigma_0^{(i)} + \sigma_1^{(i)}  \right)  \nonumber \\
  & - & \sum_{i<j} \Big ( -  \sigma_0^{(i)} \sigma_0^{(j)}  - \sigma_1^{(i)} \sigma_1^{(j)}   \nonumber \\
  &&  ~~~~~~~~+ ~ \sigma_0^{(i)} \sigma_1^{(j)} 
   +  \sigma_1^{(i)} \sigma_0^{(j)} \Big ) \geq -B_c
 \end{eqnarray}
 in which the coefficients of the linear combination of single-spin and two-spin correlation functions descend from the asymptotic value of the gradient, normalized to the common absolute value of its component. Introducing the combinations
 \begin{equation}
 {\cal S}_a = \sum_i  \sigma_a^{(i)} ~~~~~~~  {\cal S}_{ab} = \sum_{i \neq j} \sigma_a^{(i)} \sigma_b^{(j)}  
 \label{e.S}
 \end{equation}
 with $a, b = (0,1)$, the above Bell inequality takes the form
 \begin{equation}
 -{\cal S}_0 - {\cal S}_1 + \frac{1}{2} {\cal S}_{00} + \frac{1}{2} {\cal S}_{11} - {\cal S}_{01} \geq -B_c  
 \label{e.Tura}
 \end{equation}
 which corresponds to the many-body Bell inequality derived by Ref.~\cite{Turaetal2014}, with classical bound $B_c = 2N$. The saturation of the gradient to a finite value implies that the quantum data violate this Bell inequality: introducing the Bell operator 
 \begin{equation}
 \hat{\cal B} = -\hat{\cal S}_0 - \hat{\cal S}_1 + \frac{1}{2} \hat{\cal S}_{00} + \frac{1}{2} \hat{\cal S}_{11} - \hat{\cal S}_{01} 
 \end{equation}
 with  $\hat{\cal S}_a = \sum_i  \hat\sigma_a^{(i)}$ and  $\hat{\cal S}_{ab} = \sum_{i \neq j} \hat\sigma_a^{(i)} \hat\sigma_b^{(j)}$, one obtains $\langle \hat{\cal B} \rangle/N = -2.0849...$ on the $6\times 6$ lattice,
 $-2.127...$  on the $8\times 8$ lattice, $-2.147...$ on the $10\times 10$ lattice, etc. namely a consistent violation of the Bell inequality becoming stronger with increasing system size -- this is due to the relationship between the Bell inequality violation and squeezing \cite{Schmiedetal2016}, and to the increasing level of squeezing that larger lattices exhibit at the quantum Ising critical point \cite{FrerotR2018}. The quantum data are obtained via quantum Monte Carlo simulations \cite{FrerotR2018} at temperatures sufficiently low to eliminate thermal effects. 
 
 Our result is therefore two-fold: 1) the one-spin and two-spin correlation functions of the quantum critical point of the 2$d$ quantum Ising model on the square lattice violate the Bell inequality of Ref.~\cite{Turaetal2014}, Eq.~\eqref{e.Tura}; but also 2) among all the Bell inequalities that the quantum data at hand could violate,  Eq.~\eqref{e.Tura} is the one that is most strongly violated, corresponding to the facet of the local polytope which is closest to the quantum data. This is a remarkable fact, given that the inequality of Eq.~\eqref{e.Tura} is symmetric under permutation of all sites, while the quantum data are \emph{not} -- indeed the correlation functions 
  $C_{xx}^{(ij)}$ and $C_{yy}^{(ij)}$ have a strong spatial decay, which is a hallmark of quantum criticality.  This result vindicates the choice of Ref.~\cite{Turaetal2014} to restrict the search of Bell inequalities to permutationally symmetric ones, as the latter appear to be highly relevant even to quantum data which do not share the same symmetry. The same conclusion can be drawn from the example discussed in the following sections.

\subsection{Distilling the Bell inequality out of the numerical data}

The above example gives us a prescription on how to extract the violated Bell inequality out of the numerical data for the gradient of the cost function. In practice such data are noisy (as clearly shown in Fig.~\ref{Fig2_SM}): yet, after convergence is reached, a partial average of the gradient components over the last iteration steps strongly reduces the noise, and makes the pattern of the gradient components more explicit. In order to obtain relevant Bell inequalities that transcend the numerical uncertainty of the variational search, it is completely acceptable to guess some regularity in the coefficients of the inequality: once the inequality is reconstructed and the classical bound is obtained, the only metric of success of the whole operation is the violation of the inequality by the quantum data. This is also the procedure that led us to the results described in the sections that follow. 
 
\subsection{Role of the precision of the quantum data} 
 
 The example offered here, and in particular the data in Fig.~\ref{Fig2_SM}, clearly points at the role of the \emph{precision} in the quantum data in the detection of Bell non-locality. The deviation of the LV predictions from the quantum data is clearly rather small ($\sim 10^{-3}$): this means that quantum data on the correlation function with a precision significantly lower than this value are required to successfully detect non-locality. If the error on the quantum data is larger than their deviation from the LV predictions, this geometrically implies that the point representing the quantum data -- see Fig.~1 of the main text -- acquires such a width that it overlaps with the polytope of LV models, even though the exact quantum data would be outside of the polytope. In this case the quantum data analyzed in a fully unbiased way, namely compared \emph{element by element} with the predictions of LV theories, do not possess a precision sufficient to ascertain their non-local nature.  
 
 Yet, the lesson of the example coming from the 2d quantum Ising model at criticality is that Bell non-locality is optimally detected via the violation of a Bell inequality that does not treat all the elements of the magnetization profile and of the correlation functions independently, but that rather symmetrizes over all sites and pairs of sites, relying only on the combinations ${\cal S}_a$ and ${\cal S}_{ab}$. Following the spirit of Ref.~\cite{Turaetal2014}, we could have chosen to work from the start with the coarse-grain quantum data ${\cal S}_a$ and  ${\cal S}_{ab}$, and try to reproduce them with a permutationally-invariant LV theory. In this case, the symmetrized quantum data have a relative precision which is improved by a factor ${\cal O}(\sqrt{N})$ with respect to that of the individual entries of the correlation function -- assuming statistical independence between measurements of correlations at different distances  -- if the system is translationally invariant (namely if the correlation function had already been averaged over translations, thereby reducing its error by a factor ${\cal O}(\sqrt{N})$); or even by a factor ${\cal O}(N)$ in the absence of translational invariance. Working with macroscopic observables (collective spin measurements) such as ${\cal S}_a$ and  ${\cal S}_{aa}$ -- along with non-locality witnessing as opposed to strict non-locality detection -- was certainly a key ingredient toward the success of recent experiments  probing Bell non-locality in a system of $N\sim500$ (Ref.~\cite{Schmiedetal2016}) and $N\sim 5*10^5$ (Ref.~\cite{Engelsenetal2017}) quantum spins. 
 
 In conclusion, in the presence of high-precision quantum data the best strategy is to use microscopic observables (\emph{e.g.} the values of the local magnetization, of each of the two-point correlation function, etc.) so as to fully exploit the information contained in the data in search of the optimal Bell inequality. On the other hand, if this strategy fails to detect quantum non-locality because of the limited precision, one could proceed by grouping microscopic observables into mesoscopic/macroscopic observables with a reduced uncertainty. Geometrically, this amounts to project the high-dimensional quantum data point, together with the polytope of LV models, onto a lower-dimensional subspace; and to assess whether or not the projection of the quantum data falls inside or outside of the ``shadow'' of the polytope. This latter strategy has a chance to be successful, knowing that optimal Bell inequalities violated by structured quantum data may in fact be blind to their structure -- as in the example of the 2d quantum Ising model, and in the following example on the Heisenberg antiferromagnet.

\section{Proof of the many-body Pearle-Braunstein-Caves (PBC) inequality}
\label{s.PBC}

We consider a $(N,k, 2)$ Bell scenario with $k \ge 3$, where each of the $N$ parties has $k$ possible measurement settings. 
Introducing the quantities ${\cal S}_{ab}$  as in equation Eq.~\eqref{e.S}, we will prove the following Bell inequality: 
\begin{equation}
	{\cal B} = \sum_{a=0}^{k-1} {\cal S}_{aa} + \sum_{a=0}^{k-2} {\cal S}_{a,a+1} - {\cal S}_{k-1,0}  \ge -2N(k-1) := -B_c ~,
	\label{eq_BI_k}
\end{equation}
where $-B_c$ is the classical bound. We define the collective variables $X_a = \sum_{i=1}^N \sigma_a^{(i)}$, so that ${\cal S}_{ab} = X_a X_b - \sum_{i=1}^N \sigma_a^{(i)} \sigma_b^{(i)}$. We then define the quantities 
\begin{eqnarray}
	A &=& \sum_{a=0}^{k-1} X_a^2 + \sum_{a=0}^{k-2} X_a X_{a+1} - X_{k-1} X_0 ~, \nonumber \\
	B & = & \sum_{i=1}^N \left[\sum_{a=0}^{k-2} \sigma_a^{(i)} \sigma^{(i)}_{a+1} - \sigma^{(i)}_{k-1} \sigma^{(i)}_0 \right] ~,
\end{eqnarray}
so that:
\begin{equation}
	{\cal B} = A - B - kN ~.
\end{equation}
For all configurations of the variables $\sigma_a^{(i)} =\pm 1$, we have $B \le N(k-2)$, so that $-B-kN \ge -2N(k-1)$. In order to prove Eq.~\eqref{eq_BI_k} it is therefore enough to prove that $A \ge 0$. We introduce the notation $\bm u^T = (X_0, X_1, \cdots X_{k-1})$, so that $A = \bm u^T M \bm u$, with $M = \mathds{1} + M'/2$, where $M'$ is the symmetric matrix (in the Dirac notation for vectors):
\begin{equation}
	M' = \sum_{a=0}^{k-2} |a\rangle \langle a+1| - |k-1\rangle \langle 0 | + {\rm h.c.} ~.
	\label{eq_matrix_Mprime}
\end{equation} 	
In order to prove that $A \ge 0$, it is enough to prove that the eigenvalues $\epsilon_q$ ($q=0,...,k-1$) of $M'$ satisfy $\epsilon_q \ge -2$, so that $\bm u^T M \bm u \ge 0$ for any $\bm u$. The diagonalization of $M'$ is achieved in two steps: 1) a local phase transformation $|\tilde a \rangle = e^{i\pi a / k} |a\rangle$; and 2) a Fourier transformation $|\psi_q\rangle = (1 / \sqrt{k})\sum_{a=0}^{k-1} e^{2i\pi a q / k} |\tilde a \rangle$. The eigenvalues of $M'$ are $\epsilon_q = 2\cos[\pi(2q + 1) / k] \ge -2$, completing the proof that $A \ge 0$. The bound is tight for $N$ even, as $A=0$ is achieved with configurations $\sigma_a^{(i)}$ such that $X_a=0$ for all $a$, namely, with $\sigma_a^{(i)}=1$ for $i=1, \cdots N/2$ and $\sigma_a^{(i)}=-1$ for $i=N/2 + 1, \cdots N$.

\section{Violation of the many-body PBC inequality by quantum data} 
\label{s.PBCviolation}

 \begin{figure}
\includegraphics[width=0.9\columnwidth]{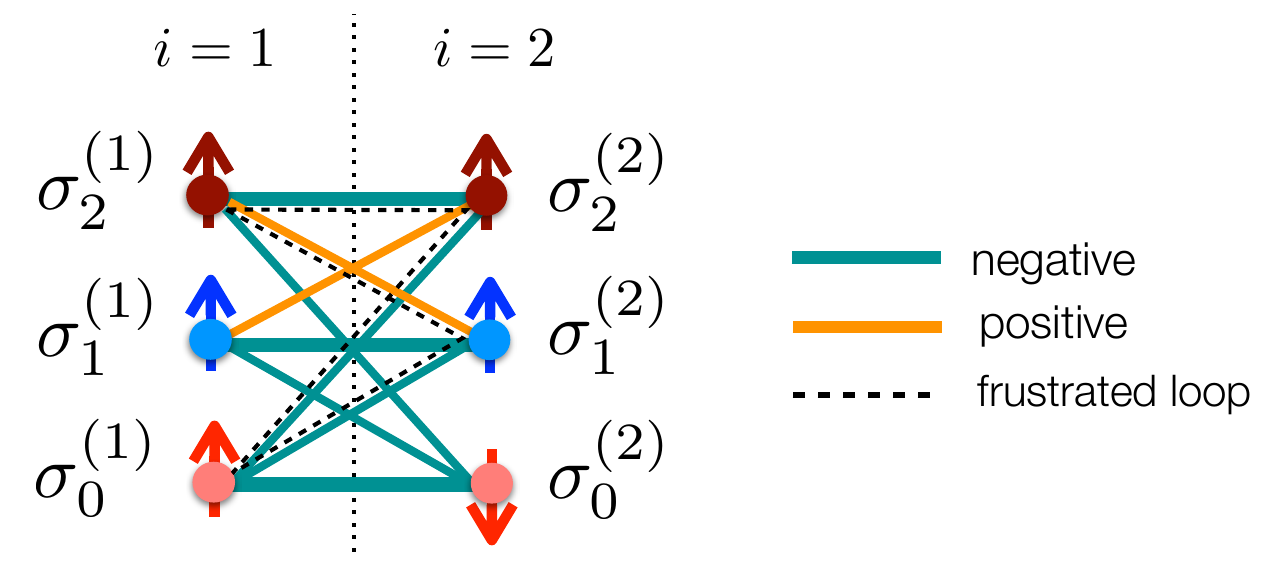}
\caption{Correlation pattern in a spin singlet (after $\pi$ rotation around the $z$ axis of \emph{e.g.} qubit 1) for the local measurements indicated in Eq.~\eqref{e.k3}. Orange (green) lines indicate positive (negative) correlations present in the quantum state. The dashed line marks one out of the four frustrated loops of correlations present in the quantum data.}
\label{Fig3_SM}
\end{figure} 

\subsection{How quantum antiferromagnetism achieves frustration of local-variable theories for $k\geq 3$}

 Before diving into the formal demonstration of the violation of the Bell inequality in Eq.~\eqref{eq_BI_k} by quantum data, it is useful to have a grasp of the physical reason behind this violation, which is also the guiding principle that allows one to identify a relevant measurement strategy on the quantum data that may lead to unveil non-locality. 

 Here we shall focus on the simple case of a singlet state shared by two qubits, and on the Bell scenario $(2,3,2)$ in which each qubit is subject to the following $k=3$ measurements 
 \begin{eqnarray}
 \hat \sigma_0 & = & \hat \sigma_x \nonumber \\
 \hat \sigma_1 & = & \cos\theta~ \hat \sigma_x + \sin\theta~ \hat \sigma_y \nonumber \\
 \hat \sigma_2 & = & \cos(2 \theta)~ \hat \sigma_x + \sin(2 \theta)~ \hat \sigma_y  ~.
 \label{e.k3}
 \end{eqnarray}
 
  The quantum data of interest are represented by the correlation functions among all possible measurements on the two qubits, $C_{ab} = \langle \sigma_a^{(1)} \sigma_b^{(2)} \rangle_{\rm Q}$, which take values:
\begin{eqnarray}
C_{00} = C_{11} = C_{22} & = & -1 \nonumber \\
C_{01} = C_{10}  = C_{12} = C_{21} & = &  - \cos\theta \nonumber \\
C_{02} = C_{20} & = & - \cos(2 \theta) \nonumber 
\label{e.pattern}
\end{eqnarray}
Choosing (as we shall do in the following) $\theta = \pi/3$ produces the simple correlation pattern $C_{00} = C_{11} = C_{22} = 2C_{01} = 2C_{12} = - 2C_{02} = -1$. This correlation pattern is pictured in Fig.~\ref{Fig3_SM}: it exhibits four frustrated correlation loops -- characterized by the presence of an odd number of correlations of the same sign -- such as in the loop $\sigma_0^{(1)} \to \sigma_1^{(2)} \to \sigma_2^{(1)} \to \sigma_2^{(2)} \to   \sigma_0^{(1)}$. This implies that the effective Hamiltonian for an LV theory capable of reproducing (or at least approaching) this correlation pattern between antiferromagnetically correlated spins must necessarily be frustrated. In fact the very correlation pattern of Eq.~\eqref{e.pattern} for two spins and $\theta = \pi/3$ is already impossible to reproduce with an LV theory. This aspect is at the heart of the violation of the Pearle-Braunstein-Caves inequality \cite{Pearle1970, BraunsteinC1990} for two qubits. 

\subsection{Coplanar measurement axes}

To test the violation of the Bell inequality Eq.~\eqref{eq_BI_k}  by a quantum state, we introduce the Bell operator for $N$ spins 
 \begin{equation}
\hat{\cal B} = \sum_{a=0}^{k-1} \hat {\cal S}_{aa} + \sum_{a=0}^{k-2} \hat {\cal S}_{a,a+1} - \hat {\cal S}_{k-1,0} ~.
\label{e.Bell_op}
\end{equation}
 where $\hat {\cal S}_{ab} =  \sum_{i\neq j}  \hat\sigma_i^{(a)} \hat\sigma_j^{(b)}$ ~.  

 In order to achieve the strongest violation of the above Bell inequality in the $(N,k,2)$ scenario by a set of quantum data, one should in principle optimize freely over all $k$ measurement axes. Pursuing this exhaustive search would lead to extremizing a complex function of $2k$ angles specifying the orientations of all measurement axes. While this is entirely feasible, we opt for a more practical scheme by which we restrict the search to the situation in which all measurement axes are co-planar and form an angle $\theta$ between consecutive axes; in other words, the measurement operators correspond to 
 \begin{equation}
 \hat{\sigma}_a = \cos(\theta a) \hat \sigma_x + \sin(\theta a) \hat \sigma_y ~~~~ (a= 0,...,k-1)
 \end{equation}  
 generalizing the scheme proposed in Eq.~\eqref{e.k3}. 
  Without loss of generality, we choose the $xy$ plane as the measurement plane and $\hat\sigma_0 = \hat\sigma_x$.
 
 The above choice of measurement axes is rather natural: consecutive measurement axes are coupled pairwise in the same way in the Bell operator $\hat{\cal B}$ (except for the last and the first one), and therefore they should form the same angles with each other, not necessarily in the same plane. The choice of coplanarity is motivated by the attempt to maximally exploit the correlations present in the the potential quantum states: indeed, having fixed the orientations of the first ($0-$th) and last ($(k-1)-$th) measurement axes, which define the plane of interest, the other measurement axes are maximally close to each other (and therefore give rise to maximally correlated measurements) if they are chosen to lie in that same plane.
  
  \subsection{Expectation value of the Bell operator}
  
 Having parametrized the choice of the measurement axes uniquely by the angle $\theta$, our goal is then to minimize the function $\langle \hat {\cal B} \rangle(\theta)$ when calculating it for a specific quantum state. Following the treatment of the Bell inequality in Sec.~\ref{s.PBC}, we rewrite the $\hat {\cal S}$ operators in terms of the collective spin operators $\hat J_a = \sum_i \hat S_a^{(i)}$ as
 \begin{equation}
 \hat {\cal S}_{ab} = 2 \left ( \hat{J}_a \hat{J}_b + \hat{J}_b \hat{J}_a \right) - \frac{1}{2} \sum_i \left ( \hat \sigma_a^{(i)} \hat \sigma_b^{(i)} + \hat \sigma_b^{(i)} \hat \sigma_a^{(i)} \right )~.
 \end{equation}
As a consequence, the Bell operator can be rewritten as 
\begin{equation}
	\hat {\cal B} = \hat{A} - \hat{B} - kN 
\end{equation} 
where 
\begin{eqnarray}
\hat{A} & = & 4 \sum_{a=0}^{k-1} \hat{J}_a^2 + 2 \sum_{a=0}^{k-2} \left( \hat{J}_a \hat{J}_{a+1} + \hat{J}_{a+1} \hat{J}_a \right ) \nonumber \\
&& - 2 \left ( \hat{J}_{k-1} \hat{J}_0 + \hat{J}_{0} \hat{J}_{k-1} \right) 
\end{eqnarray}
and 
\begin{eqnarray}
\hat{B} & = &  \frac{1}{2} \sum_{a=0}^{k-2} \sum_{i=1}^N \left( \hat{\sigma}_a^{(i)} \hat{\sigma}_{a+1}^{(i)} + \hat{\sigma}_{a+1}^{(i)} \hat{\sigma}_a^{(i)} \right ) \nonumber \\
&& - \frac{1}{2} \left ( \hat{\sigma}_{k-1}^{(i)} \hat{\sigma}_0^{(i)} + \hat{\sigma}_0^{(i)} \hat{\sigma}_{k-1}^{(i)} \right)~
\end{eqnarray}
where we have used the fact that $\hat\sigma_a^2 = \mathbb{1}$.
Given that $\hat J_a = \cos(\theta a) \hat J_x + \sin(\theta a) \hat J_y$, straightforward algebra leads to rewriting the components of the Bell operator in terms of the operators $ \hat J_x^2 $, $\hat J_y^2 $ and $\hat J_x \hat J_y + \hat J_y \hat J_x$ as follows: 
\begin{equation}
\hat A = F_x(\theta)  \hat J_x^2  + F_y(\theta) \hat J_y^2 + F_{xy}(\theta) \left (\hat{J}_x \hat{J}_y + \hat{J}_y \hat{J}_x \right)  
\end{equation}
with 
\begin{eqnarray}
F_x(\theta) & = & 2k + 2(k-1)\cos(\theta) - 4\cos[(k-1)\theta] \nonumber \\
                      && + 1 + \frac{1}{\sin\theta} \left \{ \sin[(2k-1)\theta] + \sin[2(k-1)\theta] \right \} \nonumber \\ 
F_y(\theta) & = & 2k + 2(k-1)\cos(\theta)  \nonumber \\
                      && -1 - \frac{1}{\sin\theta} \left \{ \sin[(2k-1)\theta] + \sin[2(k-1)\theta] \right \} \nonumber \\ 
F_{xy}(\theta) & = & - 2 \sin [(k-1)\theta] + \frac{1}{\sin\theta} \Big \{ 1+ \cos\theta  \nonumber \\
 && - \cos[(2k-1)\theta] - \cos[2(k-1)\theta]  \Big \}
\end{eqnarray}

A similar calculation leads to the explicit form of the $\hat{B}$ operator, which turns out to be simply proportional to the identity, $\hat B = G(\theta) N$, with 
\begin{equation}
G(\theta) =  (k-1) \cos\theta - \cos[(k-1)\theta] 
\end{equation} 

Therefore the angle $\theta$ can be found by minimizing the expectation value of the Bell operator on the quantum state of interest 
\begin{eqnarray}
\langle \hat{\cal B} \rangle (\theta) & = &  F_x(\theta)  \langle \hat J_x^2 \rangle  + F_y(\theta) \langle \hat J_y^2 \rangle \nonumber \\
&+& F_{xy}(\theta) \langle \hat{J}_x \hat{J}_y + \hat{J}_y \hat{J}_x  \rangle
- N (G(\theta) +k)~
\end{eqnarray}
in order to satisfy the condition (Bell inequality violation) $\langle \hat{\cal B} \rangle < - 2N(k-1)$. 

The above formula is completely general and it relies uniquely on the choice of coplanar measurement axes, forming equal angles between consecutive axes. In the following we shall specialize it to $U(1)$ symmetric states. 

\subsection{Bell operator for $U(1)$-symmetric states}

In the case of $U(1)$-symmetric states, invariant under a rotation around the $z$ axis, one has that  $\langle \hat{J}_x \hat{J}_y + \hat{J}_y \hat{J}_x  \rangle = 0$ and $\langle \hat J_x^2 \rangle  = \langle \hat J_y^2 \rangle$, so that the expectation value of the Bell operator reduces to 
\begin{equation} 
\langle \hat{\cal B} \rangle (\theta) = F(\theta) \langle \hat J_x^2 \rangle - N (G(\theta) +k)
\label{eq_quantum_value}
\end{equation}
where
\begin{eqnarray}
F(\theta) &=& F_x(\theta) + F_y(\theta) \nonumber \\
& =& 4\left [ k + (k-1)\cos\theta - \cos[(k-1)\theta \right ] \nonumber \\
&=& 4[k + G(\theta)] 
\end{eqnarray}
The violation of the Bell inequality is then achieved under the condition
\begin{equation}
 \frac{\langle \hat J_x^2 \rangle}{N} <  \frac{1}{4}- \frac{2(k-1)}{F(\theta)}~.
\end{equation}
The strongest violation of the Bell inequality can therefore be achieved upon maximizing the right-hand side, which implies maximising the $F$ function since $k \geq 3$. This maximum is achieved for $\theta= \pi/k$, and it takes the value
 \begin{equation}
 F_{\rm max} = F(\pi/k) = 4k\left ( 1 + \cos\frac{\pi}{k} \right )~.
 \end{equation}
 Hence the optimal witness condition for Bell non-locality of $U(1)$ symmetric states takes the form:
 \begin{equation}
 \frac{\langle \hat J_x^2 \rangle}{N} <  \beta_k = \frac{2- k + k\cos \frac{\pi}{k} }{4k \left ( 1+ \cos \frac{\pi}{k} \right)}~.
 \label{ineq_witness}
 \end{equation}
 We find the following bounds: $\beta_3 = 1 / 36 = 0.027777\cdots$, $\beta_4 = 1 / (16 + 12\sqrt{2}) = 0.030330\cdots$, $\beta_5 =  0.028885$, etc.
  The largest bound for the witness inequality Eq.~\eqref{ineq_witness} is therefore obtained for $k=4$. This is the inequality we used in the main text to probe non-locality in the low-temperature states of Heisenberg antiferromagnets.  

\vspace{.2cm}

\section{Maximal quantum violation of the PBC inequality}
\label{e.maxquantum}

In the measurement setting we have described, the maximal violation of Eq.~\eqref{eq_BI_k} is obtained when $\langle \hat J_x^2 \rangle = 0$ in a $U(1)$ symmetric state -- implying that one is actually dealing with a global singlet with higher ($SU(2)$) symmetry. The above conditions yields $\langle \hat{\cal B} \rangle = -Nk[1 +\cos(\pi /k)] < -B_c$ [see Eq.~\eqref{eq_quantum_value}]. We now prove that $-B_q := -Nk[1 + \cos(\pi / k)]$ is the maximal violation of the Bell inequality Eq.~\eqref{eq_BI_k} allowed by quantum mechanics. Specifically, we will show that 
\begin{equation}
	\hat {\cal B} + B_q \mathbb{1} \ge 0 ~,
\end{equation} 
namely that all eigenvalues of $\hat {\cal B} + B_q \mathbb{1}$ are non-negative, by rewriting this operator as a sum of squares. To define the Bell operator $\hat {\cal B}$, we first introduce the operators $\hat X_a = \sum_{i=1}^N \hat \sigma_a^{(i)}$, and $\hat {\cal S}_{ab} = (1/2)\{\hat X_a, \hat X_b\} - (1/2)\sum_{i=1}^N \{ \hat\sigma_a^{(i)}, \hat\sigma_b^{(i)} \}$, with the anti-commutator $\{\hat A, \hat B\} = \hat A \hat B + \hat B \hat A$. The only property of the $\hat \sigma_a^{(i)}$ operators that we require is that they be hermitians, and that $(\hat \sigma_a^{(i)})^2 = \mathbb{1}$, namely, that their measurement outputs $\pm 1$, regardless of the number of dimensions of the local Hilbert space upon which they act.
 
The Bell operator is then defined as in Eq.~\eqref{e.Bell_op}, yet with a generalized definition for the $\hat \sigma_a^{(i)}$ operators entering into the $\hat{\cal S}_{ab}$ operators. 
The Bell operator is a quadratic form in the $\hat \sigma_a^{(i)}$ operators. We make this form explicit by introducing the vectors of operators $\hat{\bm X}^T = (\hat X_0, \cdots\hat X_{k-1})$ and $(\hat{\bm \sigma}^{(i)})^T = (\hat \sigma^{(i)}_0, \cdots \hat \sigma^{(i)}_{k-1})$. Using the matrix $M'$ introduced in Eq.~\eqref{eq_matrix_Mprime}, we have:
\begin{eqnarray}
	\hat {\cal B} + B_q \mathbb{1} = \hat{\bm X}^T \left[\mathds{1} + \frac{1}{2}M'\right] \hat{\bm X} \nonumber \\
	+ \sum_{i=1}^N (\hat{\bm \sigma}^{(i)})^T \left[
	\left( \frac{B_q}{Nk} - 1\right) \mathds{1}
	- \frac{1}{2} M'
	\right] \hat{\bm \sigma}^{(i)} ~,
\end{eqnarray}
where we use the fact that $(\hat{\bm \sigma}^{(i)})^T \cdot \hat{\bm \sigma}^{(i)} = \sum_{a=0}^{k-1} (\sigma_a^{(i)})^2 = k~\mathbb{1}$. The symmetric matrix $M'$ has been diagonalized in Sec.~\ref{s.PBC}, so that $U^\dagger M' U = {\rm diag}(\epsilon_0, \cdots \epsilon_{k-1})$ with $\epsilon_q = 2\cos[\pi(2q+1)/k]$. Introducing the operators ${\hat{\bm Y}} =  U^\dagger \hat{\bm X} $ and $\hat{\bm \tau}^{(i)} = U^\dagger \hat{\bm \sigma}^{(i)}$, we obtain:
\begin{eqnarray}
	\hat {\cal B} + B_q \mathbb{1} = \sum_{a=0}^{k-1} \left[
		1 + \frac{\epsilon_a}{2} \right] {\hat{ Y}}_a^\dagger {\hat{ Y}}_a  \nonumber \\
	+ \sum_{i=1}^N \sum_{a=0}^{k-1} \left[
	\frac{B_q}{Nk} - 1 - \frac{\epsilon_a}{2}
	\right] (\hat{ \tau}_a^{(i)})^\dagger \hat{ \tau}_a^{(i)} ~.
\end{eqnarray}
We have $1 + \epsilon_a / 2 \ge 0$ for all $a$, and if we choose $B_q = Nk[1 + \max_a \epsilon_a/2] = Nk[1 + \cos(\pi / k)]$, we achieve a sum-of-squares decomposition, implying that for \emph{any} Hilbert space size, and \emph{any} wavefunction $|\Psi \rangle$, we have:
\begin{equation}
 \langle \Psi | \hat{\cal B} | \Psi \rangle \ge -B_q ~.
\end{equation}
The measurement protocol we have proposed on a many-body spin singlet, achieving $\langle {\cal B} \rangle = -B_q$, yields therefore the maximal violation of the Bell inequality Eq.~\eqref{eq_BI_k} allowed by quantum mechanics.

\end{document}